\documentclass[12pt]{article}

\usepackage{amsfonts,amssymb,amsmath}
\usepackage{color}
\usepackage{theorem,cite}
\usepackage{indentfirst}
\usepackage{textcomp}
\usepackage[dvips]{graphicx}

\numberwithin{equation}{section}

\definecolor{brique}{rgb}{.9,.2,0}
\definecolor{blvert}{rgb}{0,.8,.85}
\definecolor{vertcl}{rgb}{0,1,.7}
\newcommand\vertcl[1]{\textcolor{vertcl}{#1}}
\newcommand\blvert[1]{\textcolor{blvert}{#1}}
\newcommand\brique[1]{\textcolor{brique}{#1}}
\def\lapth{
\begin{picture}(164,70)(0,-15)\thicklines
\put(0,0){\vertcl{\rule{20pt}{4pt}}}
\put(19,1){\vertcl{\line(1,3){23}}} 
\put(20,1){\vertcl{\line(1,3){23}}} 
\put(21,1){\vertcl{\line(1,3){23}}}
\put(22,1){\vertcl{\line(1,3){23}}}
\put(45,70){\vertcl{\line(1,-3){23}}} 
\put(44,70){\vertcl{\line(1,-3){23}}} 
\put(43,70){\vertcl{\line(1,-3){23}}}
\put(42,70){\vertcl{\line(1,-3){23}}}
\put(2,24){\vertcl{\rule{120pt}{4pt}}}
\put(65,0){\vertcl{\rule{60pt}{4pt}}}
\put(5,37){\Huge{\brique{\textbf{L}}}} 
\put(62,37){\Huge{\brique{\textbf{PTh}}}}
\put(12,-8){\blvert{\rule{92pt}{3.5pt}}}
\put(24,-15){\blvert{\rule{57pt}{3.5pt}}}
\put(36,-22){\blvert{\rule{30pt}{3.5pt}}}
\end{picture}
\raisebox{35pt}{
\begin{minipage}{320pt}\begin{center}
\textbf{Laboratoire d'Annecy-leVieux de Physique
Th\'eorique}\\[4ex]
website: \texttt{http://lappweb.in2p3.fr/lapth-2005/}
\end{center}
\end{minipage}}\\
\vspace{10pt}\quad \hrulefill\\
\vspace{10pt}}

\newcommand{\nonu}{\nonumber \\}

\newcommand{\mb}[1]{\quad\mbox{#1}\quad}
\newcommand{\beq}{\begin{equation}}
\newcommand{\eeq}{\end{equation}}
\newcommand{\bea}{\begin{eqnarray}}
\newcommand{\eea}{\end{eqnarray}}
\newcommand{\beano}{\begin{eqnarray*}}
\newcommand{\eeano}{\end{eqnarray*}}

        \def\cL{{\cal L}}
\def\cM{{\cal M}}    \def\cN{{\cal N}}    
\def\cP{{\cal P}}

\def\cNb{\overline\cN}

\def\fS{{\mathfrak S}}

\def\fh{{\mathfrak h}}

\def\fm{{\mathfrak m}}
\def\fn{{\mathfrak n}}
\def\fp{{\mathfrak p}}
\def\fq{{\mathfrak q}}

\def\fu{{\mathfrak u}}

\newcommand{\bA}{{\mathbb A}}

\newcommand{\CC}{{\mathbb C}}

\newcommand{\II}{{\mathbb I}}

\newcommand{\half}{\frac{1}{2}}

\newcommand{\atopn}[2]{\genfrac{}{}{0pt}{}{#1}{#2}}
\newcommand{\spn}{\text{span}}
\newcommand{\ovw}[2]{\mbox{\raisebox{1.2ex}{${\substack{#1 \\ 
\downarrow \\ \textstyle #2}}$}}}
\newcommand{\npmq}{$\left(\atopn{\fn}{\fp}\,;\atopn{\fm}{\fq}\right)$}

\begin{document}

\pagestyle{empty}

\setcounter{page}{0}

%%%\section{Title}
\hspace{-1cm}\lapth

\vfill

\begin{center}

{\LARGE{\sffamily Including a phase in the Bethe equations \\[1.2ex]
 of the Hubbard model}}\\[1cm]

{\large V. Fomin, L. Frappat
and E. Ragoucy\footnote{
fomin@lapp.in2p3.fr, frappat@lapp.in2p3.fr, 
ragoucy@lapp.in2p3.fr}\\[.484cm] 
\textit{Laboratoire de Physique Th{\'e}orique LAPTH\\[.242cm]
Universit{\'e} de Savoie -- CNRS (UMR 5108) \\[.242cm]
 BP 110, F-74941  Annecy-le-Vieux Cedex, France. 
}}
\end{center}
\vfill

\begin{abstract}
We compute the Bethe equations of generalized Hubbard models, and 
study their thermodynamical limit. We
argue how they can be connected to the ones found in the context of 
AdS/CFT correspondence, in particular with the so-called dressing 
phase problem. 

We also show how the models can be interpreted, in condensed matter 
physics, as integrable multi-leg Hubbard models.
\end{abstract}

\vfill\vfill
% \rightline{\tt arXiv: [hep-th]}
\rightline{LAPTH-010/10}
\rightline{February 2010}

\newpage
\pagestyle{plain}

\section{Introduction}

The nowadays widely studied Hubbard model was introduced in the 
sixties
\cite{Hubbard,Gutzwiller} in relation with strongly correlated 
electron
systems (see \cite{Monto,book} and references therein for a review on 
the
Hubbard model). In two and three dimensions, the Hubbard 
models are unfortunately not solved yet and few results have been 
obtained. In contrast, the one-dimensional Hubbard model was found 
to be
integrable and its Hamiltonian was first diagonalized by means of the
coordinate Bethe Ansatz by Lieb and Wu in 1968 \cite{LiebWu}. Since 
then, there
have been numerous studies on this model. 

Originally described as a model of spin up and down electrons on a
one-dimensional lattice with hopping terms and nearest neighbors
interactions (resp. kinetic and potential terms of the corresponding
Hamiltonian), the Hubbard model has been then generalized in
different ways. A particularly interesting method is based on 
Shastry's
construction which was used to reveal the integrable structure of the
Hubbard model \cite{shastry,Akutsu}. The main idea is to couple two XX
model $R$-matrices through an interaction term depending on the 
coupling
constant of the Hubbard potential. The proof of the Yang--Baxter 
equation
for the obtained Hubbard $R$-matrix was given by Shiroishi and Wadati
\cite{shiro2}.

In this framework, a first generalization to the $gl(\fn)$ case was
proposed by Maassarani \cite{maasa2}. The extension to the 
superalgebraic
case $gl(\fn|\fm)$, mainly motivated by the appearance of the Hubbard 
model
in the context of $N=4$ Yang--Mills theories (see e.g. 
\cite{Rej:2005qt}),
was given in \cite{XX}. A general approach for deriving (super) 
Hubbard
models was developed in \cite{FFR}. The construction is based on the
decomposition of an arbitrary vector space (possibly infinite 
dimensional)
into a direct sum of two subspaces. The two corresponding orthogonal
projectors allow one to define a $R$-matrix of a universal XX 
model, and
then of a Hubbard model using Shastry's trick. The QISM approach 
ensures
the integrability of the models, and the properties of the obtained
$R$-matrices lead to local Hubbard-like Hamiltonians.

We continue here the investigations of the one-dimensional integrable
generalizations of the Hubbard model started in \cite{FFR}, where
Hubbard-like models based on $gl(\fn|\fm) \oplus gl(\fn'|\fm')$ were
introduced and the corresponding Hamiltonians explicitly constructed. 
The
Bethe Ansatz equations were fully derived in the XX case, while only
subsectors of the theory were examined in the Hubbard case, due to the
complexity of the calculations. A step forward was accomplished in
\cite{vFFR0609}, where the case $gl(\fn|\fm) \oplus gl(2)$ was 
investigated
and the corresponding BAE determined. In this paper, we give 
the
complete set of Bethe Ansatz equations for general 
Hubbard-like models. The structure of these BAEs is similar to 
the 
one of the usual Hubbard model, however with some (new) phases that 
depend 
on 
some Bethe roots that are quantized. These phases have to be compared 
to the ones introduced in the context of AdS/CFT correspondence 
\cite{phasestring,string}. 
Hence, we believe that this construction may be seen as a first step 
in the construction of the integrable model underlying the 
super-Yang--Mills theories. 

Another interesting application of 
these models relies on their possible interpretation as multi-leg 
Hubbard models in condensed matter physics. This may be also a new 
way to tackle the problem of two-dimensional Hubbard models.
\\

The plan of the paper is as follows. In section \ref{sect:models}, we
remind the construction for general  Hubbard
models and we set the basic notations used in 
the
paper. Then we present in section \ref{sec:BAE} the set of Bethe 
Ansatz
equations (BAEs) associated to these models, together with the 
energies and
momenta. In order to make the presentation clearer, 
the organization and the details of the calculations, based 
on the
coordinate Bethe Ansatz, are postponed in appendix \ref{sect:CBA}. 
The
section \ref{sect:TDL} deals with the thermodynamical limit. 
The Bethe
equations in this limit are derived and the ground state energy is
determined. Some applications 
of these models are presented in section \ref{sect:appli}. 
It is first emphasized that the 
corresponding Hamiltonians can be, after a Jordan--Wigner 
transformation, interpreted as multi-leg Hubbard models. 
Another point concerns the possible interpretation of the 
phases occuring in the considered Bethe equations in relation 
with the AdS/CFT correspondence. Finally, we conclude in section 
\ref{sec:conclu} on open problems.

\section{Description of the models \label{sect:models}}

Let us here remind the construction for general 
Hubbard-like models. For a given pair of algebras $\big(gl(\fn), 
gl(\fm)\big)$
there are roughly $\left[\frac{(\fn-1)(\fm-1)+1}{2}\right]$ 
non-trivial inequivalent
models, corresponding to the choice of a projector in each of the two
algebras. These choices are labelled by two integers $(\fp,\fq)$, with
$1\leq\fp\leq\fn$ and $1\leq\fq\leq\fm$, that correspond to the ranks 
of the projectors.

The Hamiltonian which we deal with in the calculations below is
derived from a transfer matrix obtained from the standard procedure 
used
for integrable systems.

At first, we define the $R$-matrix of an XX model based on the 
algebra 
$gl(\fn)$:
\begin{eqnarray}
R_{12}(\lambda) &=& \Sigma_{12}\,P_{12} + \Sigma_{12}\,\sin\lambda +
(\II\otimes\II-\Sigma_{12})\,P_{12}\,\cos\lambda
\in End(\CC^\fn)\otimes End(\CC^\fn)
\label{def:univRXX} \\
P_{12} &=& \sum_{i,j=1}^\fn E^{ij}_{1} \,  E^{ji}_{2}  \\ 
\Sigma_{12} &=& 
\sum_{i=1}^\fp  \sum_{j=\fp+1}^\fn \Big(E^{ii}_{1}\, E^{jj}_{2}  +  
 E^{jj}_{1} \, E^{ii}_{2}\Big),
\label{def:univSigma}
\end{eqnarray}
where $\lambda \in \CC$ is the spectral parameter and $\fp\in 
[1,\fn]$ is a
free integer parameter that defines the model. $E^{ij}_{x}$ denote the
elementary matrices (with entry 1 at row $i$ and column $j$ and 0
elsewhere) acting in the $x^{th}$ copy of $End(\CC^\fn)$. The XX-model
$R$-matrix (\ref{def:univRXX}) obeys the Yang--Baxter equation, is 
unitary
and regular.

\medskip

The definition of the (generalized) Hubbard $R$-matrix uses as basic
ingredient the $R$-matrix of the XX model (or its generalization), 
which
are coupled \textit{\`a la Shastry}, the coupling constant being 
related to
the potential $\fu$ of the Hubbard model. The two underlying XX models 
may be
based on two different algebras $gl(\fn)_{\uparrow}$ and
$gl(\fm)_{\downarrow}$ and depend on two different integers $\fp$ and 
$\fq$
\cite{FFR}. Let us introduce the corresponding sets of integers 
\begin{equation}
\cN_{\uparrow} =
\{1,2,...,\fp\}\,,\,\cNb_{\uparrow} = \{\fp+1,...,\fn \}
\mb{and}\cN_{\downarrow} = \{ 1,2,...,\fq \} \,,\, \cNb_{\downarrow} 
= \{
\fq +1,...,\fm \}\,.
\label{intro cN def}
\end{equation}

The $R$-matrix of the Hubbard model based on the pair
$(gl(\fn)\,,\fp\,;\,gl(\fm)\,,\fq)$ is given by
\begin{equation}
R^{\uparrow\downarrow}_{12}(\lambda_{1},\lambda_{2}) =
R^{\uparrow}_{12}(\lambda_{12})\,R^{\downarrow}_{12}(\lambda_{12}) +
\frac{\sin(\lambda_{12})}{\sin(\lambda'_{12})} \,\tanh(h'_{12})\,
R^{\uparrow}_{12}(\lambda'_{12})\,C_{\uparrow 1}\,
R^{\downarrow}_{12}(\lambda'_{12})\,C_{\downarrow 1}
\label{def:R-XXfus}
\end{equation}
where $\lambda_{12}=\lambda_{1}-\lambda_{2}$, 
$\lambda'_{12}=\lambda_{1}+\lambda_{2}$ and 
we introduced the diagonal matrix $C_\alpha$ ($\alpha = \uparrow$ or 
$\downarrow$):
\begin{equation}
C_\alpha = \sum_{j\in\cN_{\alpha}} E_\alpha^{jj} - 
\sum_{\bar{\jmath}\in\cNb_{\alpha}} 
E_\alpha^{\bar{\jmath}\bar{\jmath}}
\label{eq:matC}
\end{equation}

The coupling $h'_{12}=h(\lambda_{1})+h(\lambda_{2})$ is based on
the function $h(\lambda)$ such that
\begin{equation}
\sinh(2h)=\fu \sin(2\lambda) \,.
\label{eq:h-lambda}
\end{equation}

The $R$-matrix (\ref{def:R-XXfus}) is symmetric, regular and 
satisfies the unitary relation. Moreover, when the relation 
(\ref{eq:h-lambda}) holds,
the $R$-matrix (\ref{def:R-XXfus}) satisfies the Yang--Baxter 
equation:
\begin{eqnarray}
\label{eq:YBE}
R^{\uparrow\downarrow}_{12}(\lambda_{1},\lambda_{2})\, 
R^{\uparrow\downarrow}_{13}(\lambda_{1},\lambda_{3})\, 
R^{\uparrow\downarrow}_{23}(\lambda_{2},\lambda_{3}) 
&=& 
R^{\uparrow\downarrow}_{23}(\lambda_{2},\lambda_{3})\, 
R^{\uparrow\downarrow}_{13}(\lambda_{1},\lambda_{3})\, 
R^{\uparrow\downarrow}_{12}(\lambda_{1},\lambda_{2})\,.
\end{eqnarray}
Being equipped with an $R$-matrix with all required properties, we can
proceed to define the corresponding quantum integrable system, by
performing the following steps: monodromy matrix, transfer matrix and
Hamiltonian. The $L$-site monodromy matrix is given
\begin{equation}
T_{a<b_{1}\ldots b_{L}>}(\lambda) =
R^{\uparrow\downarrow}_{ab_{1}}(\lambda,0)\ldots 
R^{\uparrow\downarrow}_{ab_{L}}(\lambda,0)
\end{equation}
and its transfer matrix is the (super)trace in the auxiliary space:
\begin{equation}
t(\lambda)=tr_{a}T_{a<b_{1}\ldots b_{L}>}(\lambda)\,.
\end{equation}
Then the generalized Hubbard Hamiltonian reads
\begin{equation}
H = \frac{d}{d\lambda} \ln t(\lambda) \bigg\vert_{\lambda=0} 
= \sum_{x=1}^{L} H_{x,x+1}
\label{eq:HubHam}
\end{equation}
with 
\begin{equation}
H_{x,x+1} = (\Sigma P)_{\uparrow \; x,x+1 } + (\Sigma P)_{\downarrow 
\;x,x+1 } + \frac{\fu}{4} \, C_{\uparrow x} \, C_{\downarrow x} \,,
\label{eq:twositesham}
\end{equation}
where we have used periodic boundary conditions. 

\medskip

We consider generalized $(gl(\fn),\fp\ ;\ gl(\fm),\fq)$ models, 
containing 
thus
four different types of particles $\pi\uparrow$, $\bar\pi\uparrow$,
$\pi\downarrow$ and $\bar\pi\downarrow$, each type being `colored': 
the
$\pi \uparrow$-particles have `colors' $1 \uparrow, 2\uparrow,...,\fp
\uparrow$; the $\bar\pi \uparrow$-particles have `colors' $(\fp+1)
\uparrow,..., \fn \uparrow$, while the `colors' for $\pi \downarrow$ 
and
$\bar\pi \downarrow$-particles are $1 \downarrow,...,\fq \downarrow$ 
and
$(\fq+1) \downarrow,...,\fm \downarrow$ respectively.

The $L$-site Hamiltonian is given by 
   \begin{equation}
H = \sum_{x=1}^{L} \Big( (\Sigma 
P)_{\uparrow \;x,x+1 } 
+ (\Sigma P)_{\downarrow \; x,x+1 }  + \frac{\fu}{4} \ 
C_{\uparrow x} \, C_{\downarrow x} \Big),\;\; \mod{L}
\label{glnglm univHam}
    \end{equation}
where $\fu>0$ and 
    \begin{eqnarray}
&&(\Sigma P)_{\uparrow \; x,x+1 } = \sum_{i\in \cN_{\uparrow}} 
\sum_{j \in \overline\cN_{\uparrow}} \Big( E^{ij}_{\uparrow \; x}
E^{ji}_{\uparrow \; x+1} + E^{ji}_{\uparrow \; x}
E^{ij}_{\uparrow \; x+1} \Big)\,,
\\
&& (\Sigma P)_{\downarrow \; x,x+1 } = \sum_{i\in \cN_{\downarrow}} 
\sum_{j \in \overline\cN_{\downarrow}} \Big( E^{ij}_{\downarrow \; x}
E^{ji}_{\downarrow \; x+1} + E^{ji}_{\downarrow \; x}
E^{ij}_{\downarrow \; x+1} \Big)\,, \\
&& C_{\uparrow\; x} = \sum_{i\in \cN_{\uparrow}} E^{ii}_{\uparrow\;x} 
- 
\sum_{j \in \overline\cN_{\uparrow}} E^{jj}_{\uparrow\; x} 
\mb{;}\;\;
 C_{\downarrow\; x} = \sum_{i\in \cN_{\downarrow}} 
E^{ii}_{\downarrow\;x} - 
\sum_{j \in \overline\cN_{\downarrow}} E^{jj}_{\downarrow\; x}\,.
\end{eqnarray}
The corresponding system will be called a $(gl(\fn),\fp; 
gl(\fm),\fq)$ 
model, or a $\left(\atopn{\fn}{\fp}\,;\atopn{\fm}{\fq}\right)$-model.
 Note that the 
$\left(\atopn{\fn}{\fn-\fp}\,;\atopn{\fm}{\fm-\fq}\right)$-model 
is equivalent to the 
\npmq-model. The usual Hubbard model is the 
$\left(\atopn{2}{1};\atopn{2}{1}\right)$-model. The models introduced 
by Maassarani \cite{maasa2}
 are the $\left(\atopn{\fn}{1}\,;\atopn{\fm}{1}\right)$-ones.
Generalizations to the superalgebra case as done in \cite{FFR}
 will be noted 
$\left(\atopn{\fm|\fn}{\fp|\fq}\,;\atopn{\fm'|\fn'}{\fp'|\fq'}\right)$-models
with obvious notation. The generalization given in \cite{XX} 
corresponds to 
$\left(\atopn{\fm|\fn}{1|1}\,;\atopn{\fm'|\fn'}{1|1}\right)$-models.
The symmetry algebra of the 
$\left(\atopn{\fn}{\fp}\,;\atopn{\fm}{\fq}\right)$-model 
is a $gl(\fp)\oplus gl(\fn-\fp)\oplus gl(\fq)\oplus 
gl(\fm-\fq)$ algebra. 

Remark that one can add to the Hamiltonian chemical potentials 
$$
\sum_{x=1}^L\Big(\mu_{\downarrow} \sum_{i\in \cN_{\downarrow}} 
E^{ii}_{\downarrow\;x} + \bar\mu_{\downarrow}
\sum_{j \in \overline\cN_{\downarrow}} E^{jj}_{\downarrow\; x}+
\mu_{\uparrow} \sum_{i\in \cN_{\uparrow}} 
E^{ii}_{\uparrow\;x} + \bar\mu_{\uparrow}
\sum_{j \in \overline\cN_{\uparrow}} E^{jj}_{\uparrow\; x}
\Big)
$$
without perturbing integrability nor symmetry.

\section{Bethe equations of the \npmq-model\label{sec:BAE}}        

We present in this section the main result of the paper, namely the
Bethe Ansatz Equations of the model under consideration, see 
Hamiltonian
(\ref{glnglm univHam}). The details of the computation are postponed 
in the appendix and can be skipped in a first reading of the paper.

The spectrum of the generalized \npmq-Hubbard 
model is
\begin{equation}
E = \frac{\fu}{4}(L-2N_{\bar\pi})+2 \sum_{l\in \cM_{\bar\pi \uparrow} 
\cup 
\cM_{\bar\pi 
\downarrow}} \cos(k_{l}) 
\end{equation}
where the Bethe roots $k_{i}$ are parameters solution of the Bethe 
Ansatz
equations (see below) and $L$ is the number of sites. To present 
them, we
introduce integers $0\leq K\leq N\leq L$, $N_{\uparrow \pi}, 
N_{\uparrow 
\bar 
\pi}, N_{\downarrow  \pi}, N_{\downarrow  \bar\pi}\geq0$ such that 
$K=N_{\uparrow \pi}+ N_{\uparrow \bar 
\pi}+ N_{\downarrow  \pi}+ N_{\downarrow  \bar\pi}$, and sets
of integers
\begin{eqnarray}
\cM_{\pi_{\uparrow}} &=&
\{1,\ldots,N_{\uparrow \pi}\}\ , \quad \cM_{\bar \pi_{\uparrow}} = 
\{N_{\uparrow
\pi}+1,\ldots,N_{\uparrow \pi}+N_{\uparrow \bar \pi}\}\,, \\
\cM_{\pi_{\downarrow}} &=& \{N_{\uparrow \pi}+N_{\uparrow \bar \pi}+
1,\ldots,N_{\uparrow \pi}+N_{\uparrow \bar \pi}+ N_{\downarrow 
\pi}\}\,, \\ 
\cM_{\bar \pi_{\downarrow}} &=& \{N_{\uparrow \pi}+N_{\uparrow \bar 
\pi}+
N_{\downarrow \pi}+1,\ldots,K\};
\\
  \bA&=&\bA_{\pi \uparrow} \cup \bA_{\pi \downarrow} =
  \{a_{1},a_{2},\ldots,a_{N_{\uparrow \pi}}\} \cup \{a_{N_{\uparrow 
\pi} +
  N_{\uparrow \bar \pi}+ 1},a_{N_{\uparrow \pi} + N_{\uparrow \bar
  \pi}+2},\ldots,a_{N_{\uparrow \pi} + N_{\uparrow \bar \pi}+ 
N_{\downarrow
  \pi}}\} \qquad\
\end{eqnarray} 
The integers $a_{j}$ are such that $a_{i}\neq a_{j}$ for $i \in \cM_{\pi
\uparrow}$, $j \in \cM_{\pi \downarrow}$ and 
\begin{eqnarray} 
  1 \leq a_{1} < a_{2} <...< a_{N_{\uparrow \pi}} \leq N \quad , 
\quad 1
  \leq a_{N_{\uparrow \pi} + N_{\uparrow \bar \pi}+1} < ...< 
a_{N_{\uparrow
  \pi} + N_{\uparrow \bar \pi}+N_{\downarrow \pi}} \leq N
\end{eqnarray} 
Then, Bethe Ansatz equations are

\begin{eqnarray} 
&& e^{ik_{j}(L-N_{\uparrow \bar \pi})} = (-1)^{N_{\uparrow \pi}-1} 
e^{2 \pi i \frac{m_{ \uparrow \pi}}{N_{\uparrow \pi}}},\;\;m_{ 
\uparrow \pi}=1,...,N_{\uparrow \pi} 
\mb{for} j \in \bA_{\pi \uparrow},
\label{eq:BAE1} \\[1.2ex]
&& e^{ik_{j}(L-N_{\downarrow \bar \pi}-N_{\downarrow \fm})} = 
(-1)^{N_{\downarrow \pi}-1} e^{2 \pi i \frac{m_{\downarrow 
\pi}}{N_{\downarrow \pi}}},\;\;m_{\downarrow \pi}=1,...,N_{\downarrow 
\pi}
\mb{for} j \in \bA_{\pi \downarrow},
\label{eq:BAE2}\\[1.2ex]
&& e^{ik_{j}L}=(-1)^{N_{\downarrow \bar \pi}+N_{\downarrow \fm}-1} 
\prod_{m \in \cM_{\pi \downarrow}} e^{- i k_{a_{m}}}  
\prod_{l \in \cM_{\bar \pi \uparrow}} \frac{i \sin k_{j}+i 
\lambda_{l} + 
\frac{\fu}{4} }{i \sin k_{j}+i \lambda_{l} - \frac{\fu}{4}}
\prod_{m' \in \cM_{\bar \pi \downarrow}} b_{m'}
\;, \label{eq:BAE3}\\
&& \mb{for} j\in [1,N]\setminus\bA
\nonumber\\[1.2ex]
&& b_{l}^{N_{\downarrow \bar \pi}+N_{\downarrow \fm}} = e^{\frac{2 
\pi 
i}{N_{\downarrow \bar \pi}} \sum_{j=1}^{N_{\downarrow \bar \pi} - 
N_{\downarrow (\fm-1)}} n_{j}},\;\;l \in \cM_{\bar \pi 
\downarrow},\;\;arg(b_{l})<arg(b_{l+1})
\label{eq:BAE4}\\
&& \mb{with}
1 \leq n_{1} < ... < n_{N_{\downarrow \bar \pi}-N_{\downarrow 
(\fm-1)}} 
\leq N_{\downarrow \bar \pi}, 
\nonumber
\\[1.2ex]
&& \Lambda \, \prod_{\atopn{l\in \cM_{\bar \pi \uparrow}}{l \neq 
m}}
\frac{i\lambda_{m} - i\lambda_{l} + \frac{\fu}{2}}{i\lambda_{m} - 
i\lambda_{l} - \frac{\fu}{2}}  
= \prod_{\atopn{l = 1}{l \not\in \bA}}^{N} \frac{i \sin 
k_{l}+i\lambda_{m} + 
\frac{\fu}{4} }{i\sin k_{l}+i\lambda_{m} - \frac{\fu}{4}}, \mb{for} m  
\in \cM_{\uparrow \bar \pi} 
\label{eq:BAE5} \\
&&\Lambda = (-1)^{N - N_{\pi \uparrow} - N_{\pi \downarrow}} \prod_{m 
\in \cM_{\pi \downarrow}} e^{i k_{a_{m}}} 
\prod_{l \in \cM_{\bar \pi \downarrow}} b_{l}^{-1}  
\prod_{m' \in \cM_{\pi \uparrow}} e^{-i k_{a_{m'}}}
\, e^{\frac{2 \pi i}{N_{\uparrow \bar \pi}} \sum_{l'=1}^{N_{ \uparrow 
\bar \pi}-N_{\uparrow \fn}} \bar n_{l'}}, 
\label{eq:BAE6}\\
&& \mb{with} 1 \leq \bar n_{1} < ... < \bar n_{N_{\uparrow \bar 
\pi}-N_{\uparrow \fn}} \leq N_{\uparrow \bar \pi} \nonumber
\end{eqnarray} 

We have chosen $e^{1\uparrow}\otimes e^{1\downarrow}$ as the 
reference state at the first level of the Bethe ansatz (see appendix 
\ref{sect:CBA}), so that all states with 
$e^{2\uparrow}$,...,$e^{\fn\uparrow}$ and 
$e^{2\downarrow}$,...,$e^{\fm\downarrow}$ appear as excitations 
("particles") above the reference state.
$N$ is total number of such particles. 

The parameter $N_{\uparrow i}$ is the number of $i \uparrow$ 
particles for
$i=2,...,\fn$ and $N_{\downarrow j}$ is the number of $j \downarrow$
particles for $j=3,...,\fm$. $N_{\uparrow \pi} = \sum_{i=2}^{\fp}
N_{\uparrow i}$ and $N_{\downarrow \pi} = \sum_{i=2}^{\fq} 
N_{\downarrow i}
$ count the number of $ \pi$-particles with spin up and spin down
respectively. $N_{\uparrow \bar \pi} = \sum_{i=\fp+1}^{\fn} 
N_{\uparrow i}$
counts the number of spin up $\bar \pi$-particles, while 
$N_{\downarrow \bar
\pi} = \sum_{i=\fq+1}^{\fm-1} N_{\downarrow i}$ counts the number of 
spin
down $\bar \pi$-particles \textit{that are not of type $\fm 
\downarrow$}.
The reason for this latter choice will become clear in the following. 
In the
same way, $K=N_{\uparrow \pi}+N_{\uparrow \bar \pi}+N_{\downarrow
\pi}+N_{\downarrow \bar \pi}$ is the total number of particles that 
are not
of type $\fm \downarrow$.

For given integers $N_{\uparrow \pi}$, $N_{\uparrow \bar \pi}$, $N_{\downarrow
\pi}$ and $N_{\downarrow \bar \pi}$, the phases $b_{j}$ and the integers  $n_{j}$ and $\bar 
n_{j}$ 
correspond to the different "colors" that can have particles of a 
given type ($\uparrow \pi$, $\uparrow \bar \pi$, $\downarrow
\pi$ or $\downarrow \bar \pi$ types).

The integers $a_{j}$ (entering in the sets $\bA_{\pi \uparrow}$ and 
$\bA_{\pi \downarrow}$) define the order between momenta of the $\uparrow \pi$ 
and $\downarrow\pi$ particles. This order is preserved (up to a cyclic 
permutation) by the action of the Hamiltonian on the wavefunction. 

\newpage

\section{Thermodynamical limit \label{sect:TDL}}

The Bethe equations 
given in
section \ref{sec:BAE} differ from the (usual) Hubbard model's ones by some 
phases. In this section, we study them in more detail and look for 
their thermodynamical limit $L\to\infty$.

\subsection{Simplification of Bethe equations\label{sect:BE}}
The Bethe equations (\ref{eq:BAE1})-(\ref{eq:BAE5}) can be partly 
solved
for Bethe roots $k_{j}$ with $j \in \bA_{\pi \uparrow} \cup \bA_{\pi
\downarrow}$ and $b_{l}$ with $l \in \cM_{\bar \pi \downarrow}$:
\begin{eqnarray}
 k_{j} &=& \frac{ 2 \pi}{L-N_{\uparrow \bar \pi}} \left(  
\frac{N_{\uparrow \pi} -1 }{2} + \frac{m_{ \uparrow \pi}}{N_{\uparrow 
\pi}}   + I^{\uparrow \pi}_{j} \right),\mb{for} j \in \bA_{\pi 
\uparrow},
\\ 
&&m_{ \uparrow \pi}=1,...,N_{\uparrow \pi}\mb{and}   1 \leq 
I^{\uparrow \pi}_{1}<...<I^{\uparrow \pi}_{N_{\uparrow \pi}} \leq 
L-N_{\uparrow \bar \pi}
\nonumber \\
 k_{j} &=& \frac{2 \pi}{L-N_{\downarrow \bar \pi} - N_{\downarrow 
\fm}} \left(  \frac{N_{\downarrow \pi} -1 }{2} + \frac{m_{\downarrow 
\pi}}{N_{\downarrow \pi}} + I^{\downarrow \pi}_{j} 
\right),\mb{for} j \in \bA_{\pi \downarrow},
\\ 
&& m_{ \downarrow \pi}=1,...,N_{\downarrow \pi}\mb{and}   1 \leq 
I^{\downarrow \pi}_{1}<...<I^{\downarrow \pi}_{N_{\downarrow \pi}} 
\leq L-N_{\downarrow \bar \pi} - N_{\downarrow \fm}
\nonumber \\
 \ln b_{l} &=& \frac{2 \pi i}{N_{\downarrow \bar \pi} + 
N_{\downarrow \fm}} \left(  \sum^{N_{\downarrow \bar \pi} - 
N_{\downarrow \fm-1}}_{j=1} \frac{n_{j}}{ N_{\downarrow \bar \pi}}  + 
I^{\downarrow \bar \pi}_{l}\right) ,\quad l \in \cM_{\bar \pi 
\downarrow},
\;\;\label{sec3:BAE3}\\
&&\hspace{-4ex}
 1 \leq n_{1} < ... < n_{N_{\downarrow \bar \pi}-N_{\downarrow 
(\fm-1)}} 
\leq N_{\downarrow \bar \pi}\ \mbox{ ; }\ 1 \leq I^{\downarrow \bar 
\pi}_{1}<...< I^{\downarrow \bar \pi}_{N_{\downarrow \bar \pi}} \leq  
N_{\downarrow \bar \pi} + N_{\downarrow \fm}
\nonumber \\
 e^{ik_{j}L} &=& e^{2 \pi i \Phi} 
\prod_{l \in \cM_{\bar \pi \uparrow}} \frac{i \sin k_{j}+i 
\lambda_{l} + 
\frac{\fu}{4} }{i \sin k_{j}+i \lambda_{l} - \frac{\fu}{4}}
\;, \mb{for} j\in [1,N]\setminus\bA
\label{sec3:BAE4} \\ 
&&\hspace{-11ex}
 \prod_{\atopn{l = 1}{l \not\in \bA}}^{N} \frac{i \sin 
k_{l}+i\lambda_{m} 
+ \frac{\fu}{4} }{i\sin k_{l}+i\lambda_{m} - \frac{\fu}{4}} 
=
e^{2 \pi i \Psi}\, \prod_{\atopn{l\in \cM_{\bar \pi \uparrow}}{l \neq 
m}}
\frac{i\lambda_{m} - i\lambda_{l} + \frac{\fu}{2}}{i\lambda_{m} - 
i\lambda_{l} - 
\frac{\fu}{2}}  , \mb{for} m  \in \cM_{\bar \pi \uparrow}  
\label{sec3:BAE5}
\end{eqnarray}
where the phases $\Phi$ and $\Psi$ are defined by
\begin{eqnarray}
\Phi &\equiv& \frac{N_{\downarrow \bar \pi} + N_{\downarrow \fm} -
1 }{2} 
- \left( \frac{N_{\downarrow \pi} - 1}{2} + \frac{ m_{\downarrow \pi} 
}{ N_{\downarrow \pi}} \right) \frac{N_{\downarrow \pi}}{L - 
N_{\downarrow \bar \pi} - N_{\downarrow \fm} }  \nonumber \\
&& - \; \sum_{j=1}^{N_{\downarrow \pi}} \frac{I^{\downarrow 
\pi}_{j}}{L 
- N_{\downarrow \bar \pi} - N_{\downarrow \fm}} 
+ \sum_{j=1}^{N_{ \downarrow  \bar \pi} - N_{\downarrow(\fm-1)}} 
\frac{n_{j}}{ N_{\downarrow \bar \pi} + N_{\downarrow \fm} } + 
\sum_{j=1}^{N_{\downarrow \bar \pi}} \frac{I^{\downarrow \bar 
\pi}_{j}}{N_{\downarrow \bar \pi} + N_{\downarrow \fm}}  
\end{eqnarray}
and 
\begin{eqnarray}
\Psi &\equiv& \sum_{\sigma = \uparrow, \downarrow} \left( 
(-1)^{\delta_{\sigma,\downarrow} } \left( \frac{N_{\sigma \pi} - 
1}{2} + \frac{ m_{\sigma \pi} }{ N_{\sigma \pi}} \right) 
\frac{N_{\sigma \pi}}{L - N_{\sigma \bar \pi} - 
\delta_{\sigma,\downarrow} N_{\downarrow \fm} } + 
\sum_{j=1}^{N_{\sigma \pi}} \frac{I^{\sigma \pi}_{j}}{L - N_{\sigma 
\bar \pi} - \delta_{\sigma,\downarrow} N_{\downarrow \fm}} \right) + 
\nonumber \\
&& + \; \frac{N_{\uparrow \bar \pi} +  N_{\downarrow \bar \pi} + 
N_{\downarrow \fm} }{2} \; - \sum_{j=1}^{N_{ \downarrow \bar \pi} - 
N_{\downarrow \fm-1}} \frac{n_{j}}{ N_{\downarrow \bar \pi} + 
N_{\downarrow \fm} } + \sum_{j=1}^{N_{\downarrow \bar \pi}} 
\frac{I^{\downarrow \bar \pi}_{j}}{N_{\downarrow \bar \pi} + 
N_{\downarrow \fm}} + \sum_{j=1}^{N_{\uparrow \bar \pi} - N_{\uparrow 
\fn}} \frac{\bar n_{j}}{ N_{\uparrow \bar \pi}}
\end{eqnarray}
with $1 \leq \bar n_{1} < ... < \bar n_{N_{\uparrow \bar 
\pi}-N_{\uparrow 
\fn}} \leq N_{\uparrow \bar \pi}$. 

We recall that
\begin{eqnarray}
\bA=\bA_{\pi \uparrow} \cup \bA_{\pi \downarrow} = 
\{a_{1},a_{2},\ldots,a_{N_{\uparrow \pi}}\} \cup \{a_{N_{\uparrow 
\pi} + N_{\uparrow \bar \pi}+ 1},a_{N_{\uparrow \pi} + N_{\uparrow 
\bar \pi}+2},\ldots,a_{N_{\uparrow \pi} + N_{\uparrow \bar \pi}+ 
N_{\downarrow \pi}}\}
\end{eqnarray}
where the integers $a_i$ are ordered according to the inequalities
\begin{equation}
1 \leq a_{1} < a_{2} <...< a_{N_{\uparrow \pi}} \leq N \; \text{and} 
\; 1
\leq a_{N_{\uparrow \pi} + N_{\uparrow \bar \pi}+1} < a_{N_{\uparrow 
\pi} +
N_{\uparrow \bar \pi}+2} <...< a_{N_{\uparrow \pi} + N_{\uparrow \bar
\pi}+N_{\downarrow \pi}} \leq N
\end{equation}
and $a_{i}\neq a_{j}$ for $i \in \cM_{\pi \uparrow}$, $j \in \cM_{\pi
\downarrow}$.

\medskip

In order to coincide with the results in \cite{Takahashi72}, we now
slightly change the notations
\begin{equation}
\lambda_{i} \rightarrow - \Lambda_{i} \mb{and} \fu \rightarrow -4 U
\end{equation}
as well as the sign of the Hamiltonian for the energy to be equal to
\begin{equation}
E = -U(L-2N_{\bar\pi})
- 2 \sum_{l\in \cM_{\bar\pi \uparrow} \cup \cM_{\bar\pi 
\downarrow}} \cos(k_{l}) 
\end{equation}
The ground state of the model is given by the real values of the Bethe
roots $k_{j}$ and $\Lambda_{i}$. \\
Following Lieb and Wu \cite{LiebWu}, we take the logarithm of the 
Bethe
equations (\ref{sec3:BAE4}) and (\ref{sec3:BAE5}):
\begin{eqnarray}
&& k_{j}L \;\;=\;\; 2 \pi (\Phi + I_{j}) + \sum_{i=1}^{N_{\uparrow 
\bar
\pi}} \theta\Big(\frac{\Lambda_{i}-\sin k_{j}}{U}\Big) \;, \quad j \in
[1,N]\setminus \bA \\
&& \sum_{\atopn{j = 1}{j \not\in \bA}}^{N} 
\theta\Big(\frac{\Lambda_{i}-\sin
k_{j}}{U}\Big) \;\;=\;\; 2 \pi (J_{i} + \Psi) + 
\sum_{k=1}^{N_{\uparrow \bar
\pi}} \theta\Big(\frac{\Lambda_{i}-\Lambda_{k}}{2U}\Big) \;, \quad i 
\in
\cM_{\bar\pi \uparrow}
\label{Sec3:LogBae}
\end{eqnarray}
where $\theta(x)= 2 \arctan(x) \in ]-\pi,\pi]$ and we take the cut 
off for
the logarithm as $\displaystyle \frac{1}{i} 
\log\Big(\frac{x+i}{x-i}\Big) =
\pi - \theta(x) \in [0,2\pi[$.

The quantum number $I_{j}$ is integer or half-integer depending 
whether
$N_{\uparrow \bar \pi} + N_{\downarrow \bar \pi} + N_{\downarrow \fm} 
- 1 $
is even or odd, similarly $J_{i}$ is integer (half-integer) for
$N_{\uparrow \bar \pi} +1$ even (odd).

\subsection{Thermodynamical limit}
We consider the thermodynamic limit $L 
\rightarrow
\infty$ where the particle densities $\displaystyle \frac{N_{\sigma 
\bar
\pi}}{L}$, $\displaystyle \frac{N_{\sigma \pi}}{L}$ are kept fixed 
($\sigma
= \uparrow,\downarrow$). Considering the phases $\Phi$ and $\Psi$ in 
the thermodynamical limit the ratios $\displaystyle\frac{\Phi}{L}$ and
$\displaystyle\frac{\Psi}{L}$ do not vanish and depend on the 
particle densities.

% $\displaystyle\frac{\Phi}{L}$ and
%$\displaystyle\frac{\Psi}{L}$

%\begin{eqnarray}
%&& \frac{\Phi}{L} \;=\; \frac{N_{\downarrow \bar \pi} + 
% N_{\downarrow 
%\fm}}{2L} - \frac{1}{2} \frac{(N_{\downarrow \pi}/L)^2}{\left(1 - 
%\frac{N_{\downarrow \bar \pi}}{L} -  \frac{N_{\downarrow \fm}}{L} 
%\right)} \\
%&& \frac{\Psi}{L} \;=\; \frac{N_{\uparrow \bar \pi} + N_{\downarrow 
% \bar 
%\pi} + N_{\downarrow \fm}}{2L} +\frac{1}{2} \sum_{\sigma = \uparrow, 
%\downarrow} \left( 
%(-1)^{\delta_{\sigma,\downarrow} } \frac{( N_{\sigma \pi}/L )^{2}}{1 
%- \frac{N_{\sigma \bar \pi}}{L} - 
%\delta_{\sigma,\downarrow} \frac{N_{\downarrow \fm}}{L} } \right)
%\end{eqnarray}

In this limit, the real numbers $k_{j}$ and $\Lambda_{i}$ are close 
one to
each other: $k_{j+1}-k_{j} \rightarrow 0 $, $\Lambda_{i+1}-\Lambda_{i}
\rightarrow 0 $ with $L \rightarrow \infty$. They are distributed 
between
$-Q(\Phi)$ and $Q(\Phi) \leq \pi$ and $-B(\Psi)$ and $B(\Psi) < 
\infty$ for
some $Q(\Phi)$ and $B(\Psi)$. In the small intervals $dk$ and 
$d\Lambda$,
the numbers of $k_{j}$ and $\Lambda_{i}$ are $L \rho(k)dk$ and $L
\sigma(\Lambda)d\Lambda$ respectively, where $\rho(k)$ and
$\sigma(\Lambda)$ are density functions to be determined. They are
normalized as follows:
\begin{eqnarray}
\int_{-Q(\Phi)}^{Q(\Phi)} \rho(k)dk = \frac{N-N_{\uparrow 
\pi}-N_{\downarrow \pi}}{L} \quad \text{and} \quad
\int_{-B(\Psi)}^{B(\Psi)} \sigma(\Lambda)d\Lambda = \frac{N_{\uparrow 
\bar \pi}}{L}
\label{sec3:Norm}
\end{eqnarray}
The counting functions $I(k)$ and $J(\Lambda)$ are defined 
as usual from $I_{j}$ and $J_{i}$  in the continuum limit:
\begin{eqnarray}
I(k) &\!\!=\!\!& \frac{k L}{2 \pi}  - \Phi - \frac{1}{2 \pi} 
\int_{-B(\Psi)}^{B(\Psi)} d \Lambda \sigma(\Lambda) 
\theta\Big(\frac{\Lambda - \sin k }{U}\Big),
\\
J(\Lambda) &\!\!=\!\!& \frac{1}{2 \pi}\int_{-Q(\Phi)}^{Q(\Phi)}d k 
\rho(k)
\theta\Big(\frac{\Lambda - \sin k }{U}\Big) - \Psi -\frac{1}{2 \pi}
\int_{-B(\Psi)}^{B(\Psi)} d \Lambda' \sigma(\Lambda')
\theta\Big(\frac{\Lambda - \Lambda' }{2U}\Big)
\label{Sec3:countfunc}
\end{eqnarray} 
These functions are also such that $I(k_{j})=I_{j}$ and 
$J(\Lambda_{i}) =
J_{i}$. \\
Since $I(k+dk)-I(k)$ counts the number of $k$ values between $k$
and $k+dk$, we have $\displaystyle\frac{dI(k)}{dk} = L \rho(k)$, and
similarly, $\displaystyle\frac{dJ(\Lambda)}{d\Lambda} = L 
\sigma(\Lambda)$.

\medskip

\noindent Now taking the derivatives of (\ref{Sec3:LogBae}), and
considering the fact that the phases $\Phi$ and $\Psi$ do not depend 
on the
Bethe roots $k$ or $\Lambda$, we get the equations on densities, 
which are
the same as Lieb and Wu in \cite{LiebWu}:
\begin{eqnarray}
1 &\!\!=\!\!& 2\pi \rho(k) - \frac{\cos k}{U} \int_{-B}^{B} d \Lambda
\sigma(\Lambda) \theta'\Big(\frac{\Lambda -\sin k}{U}\Big) 
\\
\int_{-Q}^{Q} dk \frac{\rho(k)}{U} \theta'\Big(\frac{\Lambda - \sin
k}{U}\Big) &\!\!=\!\!& 2 \pi \sigma(\Lambda\Big) + \int_{-B}^{B} 
d\Lambda'
\frac{\sigma(\Lambda)}{2U} \theta'\Big(\frac{\Lambda - \Lambda'}{2 
U}\Big) 
\end{eqnarray}
where $k\in[-Q,Q]$, $\Lambda \in[-B,B]$ and $\theta'(x) = 
d\theta/dx(x)$.

\subsection{Ground state \label{sec:ground}}

\noindent We consider the "half-filled band" limit, defined as 
\beq
N-N_{\uparrow\pi}-N_{\downarrow \pi} = L \mb{and} 
2 N_{\uparrow \bar \pi} = N-N_{\uparrow\pi}-N_{\downarrow \pi}\,.
\eeq
Using the same arguments as in 
\cite{LiebWu}, we obtain
 $Q(\Phi)=\pi$ and $B(\Psi)=\infty$. This can be seen as follows.
Taking the normalization relations for $\rho(k)$ and $\sigma(\Lambda)$
(\ref{sec3:Norm}) and inserting there the relations between the 
counting
functions and the densities, we get:
\begin{equation}
I(Q)-I(-Q)=N-N_{\uparrow \pi}-N_{\downarrow \pi},\;\;\; J(B)-J(-B) = 
N_{\uparrow \bar \pi}
\end{equation}
Inserting the definitions of the counting functions 
(\ref{Sec3:countfunc})
in these equations, the following conditions arise: if $Q \rightarrow 
\pi$
then $N-N_{\uparrow \pi}-N_{\downarrow \pi} = L$ and if $B \rightarrow
\infty$ then $2 N_{\uparrow \bar \pi} = N-N_{\uparrow 
\pi}-N_{\downarrow
\pi}$. \\
This limit allows us to find the solution for the densities $\rho(k)$ 
and
$\sigma(\Lambda)$ by Fourier transform:
\begin{eqnarray}
\sigma_{0}(\Lambda) &\!\!=\!\!& \frac{1}{2\pi} \int_{0}^{\infty} dw \
\frac{\cos(w \Lambda)}{\cos(wU)}\;J_{0}(w) \;,\quad \Lambda 
\in[-\infty,\infty]\\
\rho_{0}(k) &\!\!=\!\!& \frac{1}{2\pi} + \frac{\cos
k}{\pi}\int_{0}^{\infty} dw\ \frac{\cos(w \sin k)
}{e^{2Uw}+1}\;J_{0}(w) \;,\quad k \in [-\pi,\pi]
\end{eqnarray}
with zeroth order Bessel function $J_{0}(x) = 
\displaystyle\frac{1}{\pi} 
\int_{0}^{\infty}dw \cos(x \sin w)$. \\
The ground state energy is then equal:
\begin{equation}
E \;=\; -U(L-2N_{\bar\pi}) - 2 \sum_{l\in \cM_{\bar\pi \uparrow} \cup
\cM_{\bar\pi \downarrow}} \cos(k_{l}) \;=\; -U(L-2N_{\bar\pi})- 4 L
\int_{0}^{\infty} dw\ \frac{J_{1}(w) J_{0}(w)}{w(e^{2Uw}+1)}
\end{equation}
with order one Bessel function $J_{1}(x) = \displaystyle\frac{x}{\pi} 
\int_{0}^{\infty}dw\ \cos(x \sin w) \cos^{2}w$.

\subsection{String hypothesis}
{From} the study of the ground state, it is tempting to conjecture 
that in the thermodynamical limit, the Bethe parameters line up 
into so-called strings, as for the usual Hubbard model. 
We remind that the string hypothesis states that all regular 
solutions $\{ k_{j} \}$ and $\Lambda_{j}$ of
Bethe equations (\ref{eq:BAE1})--(\ref{eq:BAE5}) consist of three 
kinds of
configurations:
\begin{enumerate}
\item real $k_{j} \in [-\pi , \pi]$; 
\item  $\Lambda$'s combined in
$\Lambda$-strings (of arbitrary length $n$):
\bea
 \Lambda^{n}_{a,j} &=& \Lambda^{n}_{a} + i U (n+1 - 2j)
 \,,\qquad j=1,...,n, 
\label{eq:Lstring}
\eea
\item  $2n$ $k$'s and $n$ 
$\Lambda$'s combined in $k-\Lambda$-strings:
\begin{eqnarray}
 \Lambda'^{\,n}_{a,j} &=& \Lambda'^{\,n}_{a} + i U
(n+1 - 2j)\,,\qquad j=1,...,n, \label{eq:kL-string}\\
 k^{n,j}_{a} &=& 
\begin{cases} 
\pi - \arcsin(\Lambda'^{\,n}_{a}+iU(n+1-j)), & \mb{if}
j=1,3,...,2n-1 \\
\arcsin(\Lambda'^{\,n}_{a}+iU(n-j)), & \mb{if} j=2,4,...2n-2 \\
\pi -\arcsin(\Lambda'^{\,n}_{a}-i n U), & \mb{if} j =2n
\end{cases}\nonumber
\end{eqnarray}
\end{enumerate}
The string centers $\Lambda^{n}_{a},\Lambda'^{\,n}_{a}$ are 
real, the parameter $n$ indicates the length of the string and 
\textit{a 
priori} goes
from 1 to infinity, the parameter $a$ counts the strings with the same
length: $a=1,...,M_{n}$, $M_{n}$ being the number of strings of 
length $n$.
The branch of $\arcsin(x)$ is chosen between $-\pi/2$ and $\pi/2$.
  
Using the string hypothesis inside the Bethe equations, especially for
equations $(\ref{sec3:BAE4})$ and $(\ref{sec3:BAE5})$, and taking the
logarithms, we arrive at the following form of the Bethe Ansatz 
equations
for the (real) centers of the strings, called discrete Takahashi 
equations:
\begin{eqnarray}
&& k_{j} L \;=\; 2 \pi I_{j} + 2 \pi \Phi - \sum_{n=1}^{\infty} \left(
\sum_{a=1}^{M_{n}} \theta\Big(\frac{\sin(k_{j}) - \Lambda^{n}_{a}}{n
U}\Big) + \sum_{a=1}^{M'_{n}} \theta\Big(\frac{\sin(k_{j}) -
\Lambda'^{n}_{a}}{n U}\Big) \right) 
\label{eq:taka1} \\
&& 2 L \Re(\arcsin(\Lambda'^{n}_{a} + inU))  \;=\; 2 \pi J'^{n}_{a} - 
2 \pi (\Phi + \Psi) + \sum_{\atopn{j = 1}{j \not\in \bA}}^{N-2M'} 
\theta\Big(\frac{\Lambda'^{n}_{a} - \sin{k_{j}}}{n U}\Big) + 
\nonumber \\ 
&& \hspace*{13em} + \; \sum_{m=1}^{\infty} \sum_{b=1}^{M'_{m}} 
\Theta_{nm}\Big(\frac{\Lambda'^{n}_{a} - \Lambda'^{m}_{b}}{U}\Big)
\\
&& \sum_{\atopn{j = 1}{j \not\in \bA}}^{N-2M'}
\theta\Big(\frac{\Lambda_{a}^{n} - \sin{k_{j}}}{n U}\Big) \;=\; 2 \pi
J^{n}_{a} - 2 \pi \Psi + \sum_{m=1}^{\infty} \sum_{b=1}^{M_{m}}
\Theta_{nm}\Big(\frac{\Lambda^{n}_{a} - \Lambda^{m}_{b}}{U}\Big)
\end{eqnarray}
where $j \in [1,N-2M']\setminus\bA$ in (\ref{eq:taka1}) and $M' =
\sum_{n=1}^{\infty} n M'_{n}$. The cut-off for the logarithm is taken 
as in
section \ref{sect:BE}. The function $\Theta_{nm}(x)$ is defined as:
\begin{equation}
\Theta_{nm}(x)= \sum_{l=|\frac{m-n}{2}|}^{\frac{m+n}{2}-1} \left( 
\theta\Big(\frac{x}{ 2 l } \Big) \delta_{l \neq 0} + 
\theta\Big(\frac{x}{ 2 
(l+1)} \Big) \right) 
\end{equation} 
where $\delta_{l \neq 0}$ is 0 when $l$ is 0, and 1 otherwise.

Looking at the range of the functions entering the BAEs, 
we can give bounds for the 
integers $I_{j}, J^{n}_{a},J'^{n}_{a}$:
\begin{eqnarray}
% && -\frac{L}{2} -\frac{1}{2}  
% \sum_{m=1}^{\infty}(M_{m}+M'_{m}) - \min(\Phi) \leq I_{j} 
% \leq \frac{L}{2} - \frac{1}{2} \sum_{m=1}^{\infty}(M_{m}+M'_{m}) - 
% \max(\Phi) \\
% \nonumber \\
&& -\frac{L}{2} +\frac{1}{2}  
\sum_{m=1}^{\infty}(M_{m}+M'_{m}) - \min(\Phi) \leq I_{j} 
\leq \frac{L}{2} - \frac{1}{2} \sum_{m=1}^{\infty}(M_{m}+M'_{m}) - 
\max(\Phi) \\
\nonumber \\
&& J'^{n}_{min}  \leq J'^{n}_{a} \leq J'^{n}_{max}\;\;\; \text{with}\\
% && J'^{n}_{min}=- \frac{L}{2} + \frac{N-2 M' -N_{\uparrow \pi} 
% -N_{\downarrow \pi}}{2} - \min(\Phi) - \min(\Psi) + 
% \frac{1}{2}\left(  \sum_{m=1}^{\infty} M'_{m} t_{nm} +1 \right) 
% \nonumber \\
&& J'^{n}_{min}=- \frac{1}{2}\left( {L} -N+2 M' +N_{\uparrow \pi} 
+N_{\downarrow \pi}-1-\sum_{m=1}^{\infty} M'_{m} t_{nm}  \right) 
 + \max(\Phi) + \max(\Psi) 
\nonumber \\
% && J'^{n}_{max}= \frac{L}{2} - \frac{N-2 M' -N_{\uparrow \pi} 
% -N_{\downarrow \pi}}{2} - \max(\Phi) - \max(\Psi) - 
% \frac{1}{2}\left(  \sum_{m=1}^{\infty} M'_{m} t_{nm} +1 \right) 
% \nonumber \\
&& J'^{n}_{max}= \frac{1}{2}\left( {L} -N+2 M' +N_{\uparrow \pi} 
+N_{\downarrow \pi}-1-\sum_{m=1}^{\infty} M'_{m} t_{nm} \right)
 + \min(\Phi) + \min(\Psi) 
\nonumber \\
&& J^{n}_{min}  \leq J^{n}_{a} \leq J^{n}_{max}\;\;\; \text{with}\\
% && J^{n}_{min}= - \frac{1}{2}\left( N-2 M' -N_{\uparrow \pi} 
% -N_{\downarrow \pi} - \max(\Psi) + \frac{1}{2}\left(  
% \sum_{m=1}^{\infty} M'_{m} t_{nm} +1 \right) \right) \nonumber \\
&& J^{n}_{min}= - \frac{1}{2}\left( N-2 M' -N_{\uparrow \pi} 
-N_{\downarrow \pi} - 1-\sum_{m=1}^{\infty} M'_{m} t_{nm} \right)
+ \max(\Psi)  \nonumber \\
% && J^{n}_{max}=  \frac{1}{2}\left( N-2 M' -N_{\uparrow \pi} 
% -N_{\downarrow \pi} + \min(\Psi)   -  \frac{1}{2}\left(  
% \sum_{m=1}^{\infty} M'_{m} t_{nm} +1 \right) \right) \nonumber\\
&& J^{n}_{max}=  \frac{1}{2}\left( N-2 M' -N_{\uparrow \pi} 
-N_{\downarrow \pi} - 1-
\sum_{m=1}^{\infty} M'_{m} t_{nm} \right) + \min(\Psi) \nonumber
\end{eqnarray}
with $t_{nm}= 2 \min(n,m) - \delta_{nm}$, $\min(A)$ (resp. $\max(A)$) 
is the minimal value of A and $I_{j}, J^{n}_{a},J'^{n}_{a}$ are such 
that: $I_{j}$ is integer (half-integer) if 
$\sum_{m=1}^{\infty}(M_{m}+M'_{m})$ is even (odd);  $J'^{n}_{a}$ is 
integer (half-integer) if $L - (N-2 M' -N_{\uparrow \pi} 
-N_{\downarrow \pi}) + M'_{n}$ is odd (even) and  $J^{n}_{a}$ is 
integer (half-integer) if $(N-2 M' -N_{\uparrow \pi} -N_{\downarrow 
\pi}) - \sum_{m=1}^{\infty}M_{m}t_{mn}$ is odd (even).

The minimum and maximum of $\Phi$ and $\Psi$ are:
\begin{eqnarray}
 \max(\Phi) &=& \frac{N_{\downarrow \bar \pi}+N_{\downarrow \fm} - 
1}{2} - \frac{N_{\downarrow \pi}^{2}+1}{L - N_{\downarrow \bar \pi} - 
N_{\downarrow 
\fm}} + N_{\downarrow \bar \pi } + \frac{N_{\downarrow \bar \pi} - 
N_{\downarrow (\fm-1)} (N_{\downarrow (\fm-1)}+1)/2}{N_{\downarrow 
\bar 
\pi}+N_{\downarrow \fm}} 
\nonumber \\
\min(\Phi) &=& \frac{N_{\downarrow \bar \pi}+N_{\downarrow \fm} - 
1}{2} - \frac{N_{\downarrow \pi}}{L - N_{\downarrow \bar \pi} - 
N_{\downarrow 
\fm}} - N_{\downarrow \bar \pi }  \nonumber \\
&& +\frac{(N_{\downarrow \bar \pi}-N_{\downarrow 
\fm-1})(N_{\downarrow 
\bar \pi}-N_{\downarrow (\fm-1)}+1) + N_{\downarrow \bar \pi} 
(N_{\downarrow \bar \pi}+1)}{2(N_{\downarrow \bar \pi}+N_{\downarrow 
\fm})}
\end{eqnarray}

\begin{eqnarray}
 \max(\Psi)&=& \frac{N_{\uparrow \bar \pi} + N_{\downarrow \bar
 \pi}+N_{\downarrow \fm}}{2} + (N_{\uparrow \pi}+ N_{\downarrow \pi} 
+ N_{\downarrow \bar \pi} + (N_{\uparrow \bar \pi} - N_{\uparrow 
\fn}) ) + \frac{N_{\uparrow \pi}}{L- N_{\uparrow \bar \pi}}  \nonu
&&  - \frac{ N_{\downarrow \pi} (N_{\downarrow \pi} - 1) + 
1}{L-N_{\downarrow \bar \pi} - N_{\downarrow
\fm} } - \frac{(N_{\downarrow \bar \pi} - N_{\downarrow 
\fm-1})(N_{\downarrow \bar \pi} - N_{\downarrow \fm-1}+1) - 
N_{\downarrow \bar \pi} (N_{\downarrow \bar \pi} - 
1)}{2(N_{\downarrow \bar \pi}+ N_{\downarrow \fm})} \nonu 
&& - \frac{(N_{\uparrow \bar \pi}-N_{\uparrow \fn})(N_{\uparrow \bar 
\pi}- N_{\uparrow \fn}-1)}{2 N_{\uparrow \bar \pi}}
\\ 
\min(\Psi)&=& \frac{N_{\uparrow \bar \pi} + N_{\downarrow \bar
\pi}+N_{\downarrow \fm}}{2} +  \frac{N_{\uparrow \pi}^{2} + 1}{L- 
N_{\uparrow \bar \pi}} - \frac{N_{\downarrow \fm-1}(N_{\downarrow 
\fm-1}+1)}{2(N_{\downarrow \bar \pi}+N_{\downarrow \fm})}  \nonu
&& + \frac{(N_{\uparrow \bar \pi}-N_{\uparrow \fn})(N_{\uparrow \bar 
\pi}-N_{\uparrow \fn}+1)}{2 N_{\uparrow \bar \pi}}
\end{eqnarray}
If the string hypothesis is still valid, these equations should be 
the ones to consider for the study of 
excited states above the ground state described in section 
\ref{sec:ground}. The first numerical studies we have
performed seems to indicate that when the phases $\Psi$ and $\Phi$ 
are not zero, the string hypothesis needs to be modified: the spaces 
between the imaginary parts of the Bethe roots seem not to be 
integers anymore.  However, 
more detailed numerical studies are needed to confirm it. If this 
alteration is confirmed, than the above Bethe equations will be valid 
only in the "zero phase subsector".

\section{Applications\label{sect:appli}}
\subsection{Multi-leg Hubbard models \label{sec:mutlileg}}

In this section we consider a particular 
$\left(\atopn{1|2}{1|0}\,;\atopn{1|2}{1|0}\right)$ 
Hubbard model. The Hamiltonian (\ref{glnglm univHam}) for superalgebras as introduced in \cite{FFR} 
can be written as

\beq
H = - \sum_{x=1}^{L} \Big( (\Sigma 
P)_{\uparrow \;x,x+1 } 
+ (\Sigma P)_{\downarrow \; x,x+1 }  - \frac{U}{4} \ 
C_{\uparrow x} \, C_{\downarrow x} \Big),\;\; \mod{L}
\label{eq:ham-SupAlg}
\eeq
with the gradation introduced in kinetic terms:
$$
(\Sigma P)_{\alpha\;x,x+1 } = \sum_{\bar{
\jmath}=2,3}\Big\{(-1)^{[\bar{\jmath}]}E^{1 \bar{
\jmath}}_{\alpha\;x}E^{\bar{\jmath} 1 }_{\alpha\;x+1}  +
 E^{\bar{\jmath} 1}_{\alpha\;x}
E^{1 \bar{\jmath}}_{\alpha\;x+1}\Big\},\;\;\text{with}\;\; \alpha = \uparrow,\downarrow
$$

We use the Jordan-Wigner transformation (see \cite{XX} or \cite{vFFR0609} for more details):

\bea
E^{21}_{a\,\uparrow}=(1-n^{c}_{a \, \uparrow})d^{\dagger}_{a\, \uparrow} Z^{L}_{1\,\downarrow},\;\;\;\;\;
E^{21}_{a\,\downarrow}=(1-n^{c}_{a \, \downarrow})d^{\dagger}_{a\, \downarrow} \\
E^{31}_{a\,\uparrow}=-(1-n^{d}_{a \, \uparrow})c^{\dagger}_{a\, \uparrow} Z^{L}_{1\,\downarrow},\;\;\;\;\;
E^{31}_{a\,\downarrow}=-(1-n^{d}_{a \, \downarrow})c^{\dagger}_{a\, \downarrow}
\label{JWTr}
\eea
with $n^{c}_{a\,\sigma}=c^{\dagger}_{a\,\sigma}c_{a\,\sigma}$, $n^{d}_{a\,\sigma}=d^{\dagger}_{a\,\sigma}d_{a\,\sigma}$ and $
Z^{L}_{1 \,\downarrow} = \prod_{i=1}^{L}(1-2 n^{c}_{i \, \downarrow})$.
 
The Hamiltonian (\ref{eq:ham-SupAlg}) finally can be presented in the following form:  

\bea
H &=& H^{Hub}(c,c^{\dagger},U) + H^{Hub}(d,d^{\dagger},U) + 
V_{int}(c,c^{\dagger},d,d^{\dagger},U) 
\label{eq:ham-multileg}\\
H^{Hub}(c,c^{\dagger},U)&=& - \sum_{i=1}^{L}\sum_{\sigma= 
\uparrow,\downarrow} (c^{\dagger}_{k \sigma} c_{k+1 \sigma} + 
c^{\dagger}_{k+1 \sigma} c_{k \sigma} ) + U 
\sum_{i=1}^{L}(1-2n^{c}_{k \uparrow})(1-2n^{c}_{k \downarrow}) 
\label{eq:Hub-usuel}
\eea
with

\bea
V_{int}(c,c^{\dagger},d,d^{\dagger},U)&=& - 
\sum_{i=1}^{L}\sum_{\sigma = \uparrow,\downarrow} 
\Big((c^{\dagger}_{k \sigma} c_{k+1 \sigma} + c^{\dagger}_{k+1 
\sigma} c_{k \sigma})(n^{d}_{k,\sigma}n^{d}_{k+1,\sigma}- 
n^{d}_{k,\sigma} - n^{d}_{k+1,\sigma}) 
\nonu 
&+& (d^{\dagger}_{k \sigma} d_{k+1 \sigma} + d^{\dagger}_{k+1 \sigma} 
d_{k \sigma})(n^{c}_{k,\sigma}n^{c}_{k+1,\sigma}- n^{c}_{k,\sigma} - 
n^{c}_{k+1,\sigma})\Big)   
\nonu 
&+& \frac{U}{4} \sum_{k=1}^{L}\Big( (1-2 n^{c}_{k \uparrow})(1-2 
n^{d}_{k \uparrow}) +(1-2 n^{d}_{k \uparrow})(1-2 n^{c}_{k \uparrow}) 
\nonu
&-& 2(1-n^{c}_{k \uparrow} - n^{d}_{k \uparrow})(1+2 n^{c}_{k 
\downarrow}n^{d}_{k \downarrow})  
- 2(1+2 n^{c}_{k \uparrow}n^{d}_{k \uparrow})(1-n^{c}_{k \downarrow} 
- n^{d}_{k \downarrow}) 
\nonu 
&+& (1+2 n^{c}_{k \uparrow}n^{d}_{k \uparrow})(1+2 n^{c}_{k 
\downarrow}n^{d}_{k \downarrow}) \Big)
\eea
The Hamiltonian (\ref{eq:ham-multileg}) can be interpreted as two 
periodic lines on which the electrons (described by $c,c^\dag$ for 
the first line, and by $d,d^\dag$ for the second one) 
interact via the usual Hubbard Hamitonian (\ref{eq:Hub-usuel}), plus 
a term of interaction $V_{int}$ between the two lines. Thus,
one gets a `double-row' Hubbard Hamiltonian, and considering more 
involved \npmq-models, one can construct multileg-Hubbards models.
In this way, we construct an almost two-dimensional Hubbard model that 
is still integrable.

The eigenfunctions for this Hamiltonian are made of creator operators 
$c^{\dagger}_{\uparrow}$,$c^{\dagger}_{\downarrow}$,$d^{\dagger}_{\uparrow}$ 
and $d^{\dagger}_{\downarrow}$.
They can be written in the following form and correspond to the 
solutions constructed in appendix \ref{sect:CBA}:
\bea
\Phi_{N_{1},N_{2},N_{3},N_{4}} &=& 
\sum_{\atopn{\vec x^{(1)}}{x^{(1)}_{k} \neq 
x^{(1)}_{l}}}
\sum_{\atopn{\vec x^{(2)}}{x^{(2)}_{k} \neq x^{(2)}_{l}, 
x^{(1)}_{l}}}
\sum_{\atopn{\vec x^{(3)}} {x^{(3)}_{i} \neq x^{(3)}_{j}}}
\sum_{\atopn{\vec x^{(4)}} {x^{(4)}_{i} \neq x^{(4)}_{j}, 
x^{(3)}_{l}}} 
\Psi'(\vec x^{(1)},\vec x^{(2)},\vec x^{(3)},\vec x^{(4)}) \nonu
&&
\prod_{j=1}^{N_{1}} c^{\dagger}_{x^{(1)}_{j} \uparrow} \ 
\prod_{j=1}^{N_{2}} c^{\dagger}_{x^{(2)}_{j} \downarrow} \
\prod_{j=1}^{N_{3}} d^{\dagger}_{x^{(3)}_{j} \uparrow} \
\prod_{j=1}^{N_{4}} d^{\dagger}_{x^{(4)}_{j} \downarrow} \
|0> = (-1)^{N_{1}+N_{2}} 
\phi[\bar{A}] \\
\mbox{and} && \bar{A}
=(\overbrace{3\uparrow,...,3\uparrow}^{N_{1}},
\overbrace{3\downarrow,...,3\downarrow}^{N_{2}},
\overbrace{2\uparrow,...,2\uparrow}^{N_{3}}
\overbrace{2\downarrow,...,2\downarrow}^{N_{4}})\,.
\eea
where $\phi[\bar{A}]$ is the eigenfunction given in (\ref{excited st 
glnglm}). The particles 
$c^{\dagger}_{\uparrow}$,$c^{\dagger}_{\downarrow}$, 
$d^{\dagger}_{\uparrow}$ and $d^{\dagger}_{\downarrow}$ 
correspond to $3\uparrow$, $3\downarrow$, $2\uparrow$ and 
$2\downarrow$ respectively. The particles $c^{\dagger}_{\sigma}$ and 
$d^{\dagger}_{\sigma}$ can be identified with a (spin up and 
down) electrons. 

The energy of the excited state $\Phi_{N_{1},N_{2},N_{3},N_{4}}$ reads
$$
E = U \Big(L-2N\Big) - 2 \sum_{l=1}^{N} \cos 
k_{l}\,\;\;\text{with}\;\;N=N_{1}+N_{2}+N_{3}+N_{4}
$$
where the parameters $k_{l}$ are Bethe roots defined by equations: 
\begin{eqnarray} 
&& e^{ik_{j}L}= \prod_{l = 1}^{N_{1}+N_{3}} \frac{ \lambda_{l} - \sin k_{j} - 
i U }{ \lambda_{l} - \sin k_{j} + i U}
\prod_{i =1}^{N_{4}} b_{i}
\;, \label{sec4eq:BAE1} \mb{for} j\in [1,N]
\\[1.2ex]
&& b_{i}^{N_{2}+N_{4}} = 1,\;\;arg(b_{i})<arg(b_{i+1}), \mb{with}
i \in [1,N_{4}], 
\label{sec4eq:BAE2}
\\[1.2ex]
&& \prod_{j = 1}^{N} \frac{ \lambda_{l} - \sin 
k_{j} - i U }{\lambda_{l} - \sin k_{j} + i U}  
= e^{\frac{2 \pi i}{N_{1}+N_{3}} \sum_{i=1}^{N_{3}} \bar n_{i}} 
\prod_{i=1}^{N_{4}} b_{i} \prod_{\atopn{m = 1}{m \neq 
l}}^{N_{1}+N_{3}}
\frac{\lambda_{l} - \lambda_{m} - 2 i U}{\lambda_{l} - 
\lambda_{m} + 
2 i U}, \label{sec4eq:BAE3} \\
&& \mb{with} 1 \leq \bar n_{1} < ... < \bar n_{N_{3}} \leq N_{1}+N_{3},  
\mb{for} l \in [1,N_{1}+N_{3}] \nonumber
\end{eqnarray} 

\subsection{Comparison with AdS/CFT}
The (energy) spectrum of the (usual) Hubbard model has been shown 
\cite{Rej:2005qt} to 
reproduce correctly the 
spectrum of the dilation operator in the $su(2)$ sub-sector up to 
three 
loops. The perturbation theory is done at $g\to0$ on the SYM-side, 
while it is done at $\fu\to\infty$ on the Hubbard's side.
 Starting at fourth loop, the corrections \cite{phasestring} differ with the Hubbard 
model. The string Bethe equations (for the $su(2)$ subsector) read
\beq
\left(\frac{x(u_{k}+\frac i2)}{x(u_{k}-\frac i2)}\right)^L
=\prod_{\atopn{j=1}{j\neq k}}^M 
\frac{u_{k}-u_{j}+i}{u_{k}-u_{j}-i}\,e^{i\theta(u_{k},u_{j})}\,
\mb{with}\begin{cases}
u_{j}=\half\mbox{cotan}(\half p_{j})\\
x(u)= 
\frac u2\left(1+\sqrt{1-\frac{2g^2}{u^2}}\right)\end{cases}
\label{eq:Bethestring}
\eeq
where $\theta(u_{k},u_{j})$ is the so-called dressing phase that 
takes care of the discrepency to the Hubbard model. In general, 
this phase has two different origins. One is 
just related to the asymptotic Bethe ansatz and the scattering matrix 
of the model, and the other cause comes from the so-called 
wrapping problem, occuring when the order of the expansion in $g$ is bigger 
than the length of the chain, $L$. 
When considering the $su(2)$ sector 
and the Hubbard model, the wrapping problem has been already 
`solved': the Hamiltonian is local, and it is the expansion in term 
of the coupling constant that makes appear different spin chain models 
with long range interactions. Hence, for this sector, it is only the 
dressing factor that has to be implemented.

In the Bethe equation (\ref{eq:BAE5}), such a phase occurs, 
 that depends on some Bethe parameters.
Unfortunately, for the present models, the Bethe parameters involved 
in the phase seem not to be of the type $u_{j}$. However, we believe 
that the present construction could be a first step for the 
construction of an integrable model possessing such a dependence. In 
particular, since the models we construct can be interpreted as 
multi-leg 
Hubbard models (see section \ref{sec:mutlileg}), they are free from 
any wrapping problem. In the context of SYM theories, 
one should stick to the single circle interpretation, with electrons possessing some internal 
degrees of freedom (as it is done in the Bethe ansatz of appendix 
\ref{sect:CBA}). Then, to get a true $su(2)$ model, one would have to 
integrate over the internal degrees of freedom, to get an effective 
model. 

To be more appealing, we reformulate the Bethe equations 
(\ref{sec4eq:BAE1})-(\ref{sec4eq:BAE3})  with the 
notation and half-filling constraint:
\begin{equation}
2\,U=\frac1{g\sqrt{2}} \mb{;} 
\frac{\lambda_{\ell}}{2U}=u_{\ell}
\mb{;} N=N_{1}+N_{2}+N_{3}+N_{4}=L 
\mb{;} N_{1}+N_{3}=M
\end{equation} 
Forgetting for a while the state multiplicity, eq. (\ref{sec4eq:BAE2}), we also set
\begin{equation}
\prod_{i =1}^{N_{4}} b_{i}=e^{\frac{2i\pi}{L-M}\theta_{24}} \mb{and} 
e^{\frac{2 \pi i}{N_{1}+N_{3}} \sum_{i=1}^{N_{3}} \bar n_{i}}
=e^{\frac{2i\pi}{M}\theta_{13}}
\end{equation} 
Then, we get:
\begin{eqnarray} 
 e^{ik_{j}L} &=& e^{\frac{2i\pi}{L-M}\theta_{24}}
\prod_{\ell = 1}^{M} \frac{u_{\ell} - g\sqrt{2}\,\sin k_{j} - 
\frac{i}2  }{ u_{\ell} - g\sqrt{2}\,\sin k_{j} + \frac{i}2 }
 \mb{for} j=1,\ldots,L\quad
\\[1.2ex]
 \prod_{j = 1}^{L} \frac{ u_{l} - g\sqrt{2}\,\sin 
k_{j} - \frac{i}2 }{u_{l} - g\sqrt{2}\,\sin k_{j} + \frac{i}2}  
&=& e^{\frac{2i\pi}{M}\theta_{13}+\frac{2i\pi}{L-M}\theta_{24}} 
 \prod_{\atopn{m = 1}{m \neq l}}^{M}
\frac{u_{l} - u_{m} -  i}{u_{l} - u_{m} + i}  
\mb{for} l =1,..,M
\end{eqnarray} 
The first equation can be used to determine the  the Bethe 
roots $k_{j}$, while the 
second one has to be compared with 
(\ref{eq:Bethestring}).

\section{Conclusion\label{sec:conclu}}
We have presented the Bethe equations for generalized Hubbard models, 
using the coordinate Bethe ansatz. They look similar to the usual 
Hubbard model ones, but a phase that appears in the equations. This 
phase has to be compared to the one appearing in the string Bethe 
equations. Unfortunately, the present phase is constant 
(more precisely depends only on the non dynamical parameters associated 
to the $\bar\pi\downarrow$ particles,), while the one 
appearing in the string context depends on the Bethe parameters 
(associated here to the $\bar\pi\uparrow$ particles). We 
believe our construction can be used as a first step for constructing 
an integrable model that would reproduce the expected phase. 
An open problem is thus to look for an amendment of the construction to provide $k$-dependent 
phases for AdS/CFT. In that respect, let us note the studies done in 
\cite{dolcini} that  provides an alternative way to get generalized 
Hubbard models.

The thermodynamical limit has been also
discussed and a string hypothesis presented. In some cases (zero 
phase subsectors), it seems clear that the string hypothesis is 
correct and similar to the one of the (usual) Hubbard model.
However, when the phases are not zero there are 
presently not enough data to conclude on the validity of the string 
hypothesis. Thus, we think that more
numerical studies are needed to confirm or modify the string hypothesis in the 
thermodynamical limit.

Finally, we think that our models can be used in condensed matter 
physics to define of multi-leg Hubbard models. Contrarily to the real 
two-dimensionnal Hubbard model, these multi-leg Hubbard models are still 
integrable. Hence, they can be used to get more insight on the 
two-dimensionnal model using 
integrable methods.

\newpage
\appendix

\section{Coordinate Bethe ansatz \label{sect:CBA}}

The derivation of the Bethe Ansatz equations of the model is 
based on
the use of the coordinate Bethe Ansatz. 
The diagonalization of the Hamiltonian (\ref{glnglm univHam}), which
involves $\fn+\fm$ types of particles, is done in two main steps. 
The Hamiltonian eigenfunctions are written as linear 
combinations of
plane waves, whose coefficients are found to be solutions of a 
new eigenvalue
problem, the Hamiltonian of which shows up as a chain of 
$S$-matrices. This
is the first auxiliary problem.

In order to diagonalize this auxiliary Hamiltonian, the corresponding 
wave
function is decomposed as excitations above a suitably chosen 
pseudo-vacuum
with some coefficients. Using a recursive representation of the 
auxiliary
Hamiltonian, recursive relations between these coefficients can be 
found.
The resolution of these relations depend on the effective structure 
of the
Bethe roots, since the choice of the pseudo-vacuum breaks the symmetry
between spin up and spin down particles. The different cases are then 
considered.

\subsection{Coordinate Bethe Ansatz, level one}        
    
The states corresponding to $N$ excitations are written as
    \begin{equation}
\phi[\overline{A}] = \sum_{\mathbf{x}\in [1,L]}
\Psi[\mathbf{x},\overline{A}] \prod_{i=1}^{N}
e^{A_{i}}_{x_{i}} \phi_{0} \;,
\label{excited st glnglm}
\end{equation} 
with $\bar{A}
=(\overbrace{2\uparrow,...,2\uparrow}^{N_{\uparrow 2}},,...,
\overbrace{\fn\uparrow,...,\fn\uparrow}^{N_{\uparrow \fn}},
\overbrace{2\downarrow,...,2\downarrow}^{N_{\downarrow 
2}},...,\overbrace{\fm\downarrow,...,\fm\downarrow}^{N_{\downarrow 
\fm}})$. The sum is done on $\mathbf{x} = (x_{1},x_{2},...,x_{N})$  
without
points where two particles with the same spin coincide.
The vacuum is chosen as
\begin{equation}
\phi_{0} = 
\prod^{L}_{k=1} e^{1\uparrow}_{k} e^{1\downarrow}_{k}\,.
\label{vacuum}
\end{equation}

Acting with the Hamiltonian (\ref{glnglm univHam}) on the
excited state, we get the eigenvalue equation for the $\Psi$ function:
\begin{eqnarray}
\sum_{l \in \cM_{\bar\pi \uparrow} \cup \cM_{\bar\pi 
\downarrow}} \left( \Psi[\mathbf{x} - \mathbf{e}_{l} + \delta^{-}_{1} 
\mathbf{e}_{\pi},
\overline{A}] \Delta^{-}_{l} + \Psi[\mathbf{x}
+ \mathbf{e}_{l} - \delta^{+}_{1} \mathbf{e}_{\pi} ,\overline{A}] 
\Delta^{+}_{l} \right)  +  \nonumber \\  
\Big(\frac{\fu}{4}(L-2N_{\bar\pi}) +  \fu\sum_{l \in  \cM_{\bar 
\pi_{\uparrow}}} 
\sum_{l \in \cM_{\bar \pi_{\downarrow}}} \delta(x_{l}=x_{n})
 - E
\Big)\Psi[\mathbf{x},\overline{A}]
\Delta^{3} = 0
\label{psi eq glnglm univ}
\end{eqnarray}
where $\mathbf{e}_{m}$ is an elementary vector in $\CC^N$ with $1$ on 
the
$m$ position and $0$ elsewhere. The notation $\Psi[\mathbf{x} -
\mathbf{e}_{l} + \delta^{\pm}_{1} \mathbf{e}_{\pi}]$ means that, if 
under
the Hamiltonian's action some $\bar\pi$-particle (at position $x_{l}$ 
with
some spin $\alpha_{l}$) is moved to a place already occupied by some 
other
$\pi$-particle with same spin, then they exchange their places. The symbols
$\Delta^{\pm}_{l},\Delta^{3}$ represent the exclusion principle for
$\bar\pi$-particles: they are equal to zero if two such particles with
the same spin coincide on the same site. Explicitly, they read:
\begin{eqnarray} 
\Delta^{\pm}_{m} &=& \prod_{l\neq m}
\prod_{n \neq m} \delta^{\updownarrow}(x_{l} \neq x_{n})\ \prod_{l}
\delta^{\updownarrow}(x_{l} \neq x_{m}) \delta^{\updownarrow}(x_{l} 
\neq
x_{m} \pm 1) \;, \\
\Delta^{3} &=& \prod_{l\neq n} 
\delta^{\updownarrow}(x_{l} \neq x_{n}) \;,
\end{eqnarray}
where
\begin{equation}
  \delta^{\updownarrow}(x_{l} \neq x_{n}) = 1 -
  \delta(x_{l}-x_{n}) \quad \text{for} \quad l,n  \in \cM_{\bar \pi 
  \uparrow}\;\text{or}\;\cM_{\bar \pi \downarrow}.
  \label{symbols glnglm univ}
\end{equation}
We assume the Bethe hypothesis for the general solution of 
$\Psi[\mathbf{x},\overline{A}]$.
Dividing the coordinate space $(x_{1},x_{2},..,x_{N})$ into $N!$
sectors, we write for $x_{Q(1)} < x_{Q(2)} < .. < x_{Q(N)}$,  
\begin{equation}
\Psi^{P_{\pi\bar\pi}}_{Q}[\mathbf{x},\overline{A}]=
\sum_{P'=P_{\pi}P_{\bar\pi}} \Phi_{\overline{A}}(\hat{P} Q, 
\hat{P}^{-1})
e^{i \hat{P}\mathbf{k} \mathbf{x}} \;, \quad \hat{P}=P_{\pi\bar\pi}P'
\label{Psi ansatz}
\end{equation}
where $k_{1},k_{2},..,k_{N}$ are unequal numbers (the {\itshape Bethe
roots}), $Q=[q_{1},q_{2},..,q_{N}]$ is an element of the permutation 
group
$\fS_N$ and $\hat P \mathbf{k} \mathbf{x} = \sum_{i} k_{\hat P (i)} 
x_{i}$.
We consider the permutation $\hat P$ in its 
factorized
form $\hat P = P_{\pi\bar\pi} P_{\pi} P_{\bar\pi}$, where 
$P_{\pi\bar\pi}$
is a global and fixed permutation of $\pi$ and $\bar\pi$ particles, 
while
the terms $P_{\pi}$ and $P_{\bar\pi}$ permute $\pi$ and $\bar\pi$ 
particles
separately.

The energy depends on the "global" $P_{\pi\bar\pi}$ permutation:
\begin{equation}
E^{P_{\pi\bar\pi}} = 2 \sum_{l\in \cM_{\bar\pi \uparrow} \cup 
\cM_{\bar\pi \downarrow}} \cos(k_{P_{\pi\bar\pi}(l)}) + 
\frac{\fu}{4}(L-2N_{\bar\pi})\,.
\label{free energy glnglm univ}
\end{equation}
The coefficients $\Phi_{\overline{A}}(\hat{P} Q, \hat{P}^{-1})$ in
(\ref{Psi ansatz}) are not all independent. Indeed, using 
the symmetry of the wave function and the application of the 
Hamiltonian
represented in (\ref{psi eq glnglm univ}), it is possible to reduce 
their
number in several cases:

\medskip

1. For identical particles of any kind ($\pi$ or $\bar\pi$), the wave
function satisfies the following symmetry property
\begin{equation}
\Psi^{P_{\pi\bar\pi}}_{Q}[\mathbf{x},\overline{A}] = 
\Psi^{P_{\pi\bar\pi}}_{Q \Pi_{i i+1}}[\Pi_{Q(i) 
Q(i+1)}\mathbf{x},\Pi_{Q(i) Q(i+1)} \overline{A}]
\end{equation}
that leads to the relation
\begin{equation}
\Phi_{\overline{A}}(\hat{P} Q, \hat{P}^{-1}) = 
\Phi_{\overline{A}}(\hat{P} Q, (\hat{P} \Pi_{Q(i) Q(i+1)})^{-1})
\label{WF sym glnglm univ}
\end{equation}
where $\Pi_{ab}$ is the permutation of objects $a$ and $b$, the 
indices
$Q(i),Q(i+1)$ correspond to identical particles.

\medskip

2. For particles with the same spin we also impose some kind of weak 
"exclusion
principle" such that the wave function vanishes if two particles with 
the
same spin coincide (\textit{but the particles can differ by their 
$\pi$ or
$\bar\pi$ type}). This principle is verified for $\pi$-particles if 
we use
(\ref{psi eq glnglm univ}) with two particles at positions $x_{Q(i)} =
x_{Q(i+1)} - 1$ (in the sector $x_{Q(1)} \ll .. \ll x_{Q(i)} < 
x_{Q(i+1)}
\ll x_{Q(N)}$, the notation $\ll$ means that the particles are far 
enough
from each other). We generalize this condition to any kind of 
particles. It leads to:
\begin{equation}
\Psi^{P_{\pi\bar\pi}}_{Q}[\mathbf{x},\overline 
A]_{x_{Q(i)}=x_{Q(i+1)}} = 0
\;, \quad A_{Q(i)},A_{Q(i+1)} \in \cN_{\uparrow} \cup \cN_{\downarrow}
\;\;\text{or}\;\;\overline\cN_{\uparrow} \cup 
\overline\cN_{\downarrow}
\end{equation}
that is to say
\begin{equation}
\Phi_{\overline A}(\hat{P} Q, \hat{P}^{-1}) = - \Phi_{\overline 
A}(\hat{P} Q \Pi_{i i+1}, (\hat{P} \Pi_{Q(i) Q(i+1)})^{-1})
\label{Phi'=-Phi glnglm univ}
\end{equation}
for any $i$ such that $A_{Q(i)},A_{Q(i+1)} \in \cN_{\uparrow} \cup 
\cN_{\downarrow} \;\;\text{or}\;\;\overline\cN_{\uparrow} \cup 
\overline\cN_{\downarrow}$.

\medskip

3. For particles with different spins, there is no exclusion 
principle and
they can be at the same site. Thus, let us consider the case: 
$x_{Q(1)} \ll
.... \ll x_{Q(i)} = x_{Q(i+1)} \ll x_{Q(N)}$ and ask for 
the continuity of the wave function
 on the boundary of the two domains $Q$ and $Q \Pi_{i 
i+1}$:
\begin{eqnarray} 
\Psi^{P_{\pi\bar\pi}}_{Q}[\mathbf{x}, \overline 
A]_{x_{Q(i)}=x_{Q(i+1)}} = \Psi^{P_{\pi\bar\pi}}_{Q\Pi_{i 
i+1}}[\mathbf{x},\overline A]_{x_{Q(i)}=x_{Q(i+1)}} \,.
\end{eqnarray}
For two $\bar\pi$-particles, it implies the relation, for $A_{Q(i)} 
\in
\overline \cN_{\uparrow (\downarrow)}$ and $A_{Q(i+1)} \in \overline
\cN_{\downarrow (\uparrow)}$,
\begin{eqnarray}
&& \Phi_{\overline A}(\hat{P} Q, \hat{P}^{-1}) + \Phi_{\overline 
A}(\hat{P}
Q \Pi_{i i+1}, (\hat{P}\Pi_{Q(i) Q(i+1)})^{-1}) = \nonumber \\ 
&& \qquad \qquad \qquad
\Phi_{\overline A}(\hat{P} Q , (\hat{P} \Pi_{Q(i) Q(i+1)})^{-1}) +
\Phi_{\overline A}(\hat{P} Q \Pi_{i i+1}, \hat{P}^{-1})
\label{WF cont pibar glnglm univ}
\end{eqnarray}
while for two $\pi$-particles it leads, for $A_{Q(i)} \in 
\cN_{\uparrow
(\downarrow)}$ and $A_{Q(i+1)} \in \cN_{\downarrow (\uparrow)}$, to
\begin{equation}
\Phi_{\overline A}(\hat{P} Q \Pi_{i i+1}, 
\hat{P}^{-1})=\Phi_{\overline A}(\hat{P} Q, \hat{P}^{-1})\,.
\label{WF cont pi glnglm univ}
\end{equation}
When one particle is of type $\pi$ and the other one of type 
$\bar\pi$, we
find, for $A_{Q(i)} \in \cN_{\uparrow (\downarrow)}$ and $A_{Q(i+1)} 
\in
\overline \cN_{\downarrow (\uparrow)}$,
\begin{equation}
\Phi_{\overline A}(\hat{P} Q \Pi_{i i+1}, 
\hat{P}^{-1})=\Phi_{\overline A}(\hat{P} Q, \hat{P}^{-1}) 
\,.
\label{WF cont pi pibar glnglm univ}
\end{equation}
There is another condition for $\bar\pi$-particles when using 
(\ref{psi eq
glnglm univ}). This is the usual relation on the coefficient
$\Phi_{\overline A}(\hat{P} Q, \hat{P}^{-1})$ obtained in the 
$\left(\atopn{2}{1}\,;\atopn{2}{1}\right)$ Hubbard model for spin up 
and spin down interacting electrons:
\begin{eqnarray}
\sum_{m \neq Q_{i},Q_{i+1}} \Big( 
\Psi^{P_{\pi\bar\pi}}_{Q}[\mathbf{x} -
\mathbf{e}_{m}, \overline A] + \Psi^{P_{\pi\bar\pi}}_{Q}[\mathbf{x} +
\mathbf{e}_{m}, \overline A] \Big) + \big(\frac{\fu}{4}(L-2N_{\bar\pi}) 
+ \fu - E
\big)\Psi^{P_{\pi\bar\pi}}_{Q}[\mathbf{x}, \overline A] + \nonumber \\
\Psi^{P_{\pi\bar\pi}}_{Q}[\mathbf{x} - \mathbf{e}_{Q_{i}}, \overline 
A] +
\Psi^{P_{\pi\bar\pi}}_{Q \Pi_{i i+1}}[\mathbf{x} + \mathbf{e}_{Q_{i}},
\overline A] +\Psi^{P_{\pi\bar\pi}}_{Q \Pi_{i i+1}}[\mathbf{x} -
\mathbf{e}_{Q_{i+1}}, \overline A] + 
\Psi^{P_{\pi\bar\pi}}_{Q}[\mathbf{x} +
\mathbf{e}_{Q_{i+1}}, \overline A] = 0 \nonumber 
\end{eqnarray}
Skipping the intermediate calculations and combining the results with
(\ref{WF cont pibar glnglm univ}), we can write the conditions on
$\Phi(\hat{P} Q, \hat{P}^{-1})$ in a matrix form:
\begin{equation}
 \begin{pmatrix}
  \Phi(\Pi_{ab}\hat{P} Q, \hat{P}^{-1}) \\
  \Phi(\Pi_{ab} \hat{P} Q, (\Pi_{ab} \hat{P})^{-1})
  \end{pmatrix} =
\begin{pmatrix}
  t_{ab} & r_{ab}  \\
  r_{ab} & t_{ab}
  \end{pmatrix}
\begin{pmatrix}
  \Phi(\hat{P} Q, \hat{P}^{-1}) \\
  \Phi(\hat{P} Q, (\Pi_{ab} \hat{P})^{-1})  
\end{pmatrix}
  \label{Phi=MPhi glnglm univ}
\end{equation}
with $a=\hat{P}Q(i)$, $b=\hat{P}Q(i+1)$ and  
\begin{equation}
t_{ab} = \frac{2i(\lambda_{a} - \lambda_{b} )}{\fu + 2i (\lambda_{a} - 
\lambda_{b}) }, \; \; r_{ab} = \frac{ - \fu }{\fu + 2i
(\lambda_{a} - \lambda_{b}) },\;\;\lambda_{a} = \sin k_{a}
  \label{r,t glnglm univ}
\end{equation}
These equations hold for any type of $\bar\pi$-excitations (any
value of $A_{Q_{i}}$,$A_{Q_{i+1}}$ being $\bar\pi$-particles). 

\medskip

4. Finally we consider the interaction between $\pi$ and $\bar\pi$ 
particles with the same spin. Let $A_{Q(i)}$ be a $\bar\pi$-particle, 
$A_{Q(i+1)}$ be a $\pi$-particle, with coordinates  
$x_{Q(i)}=x_{Q(i+1)}-1$, thus $\delta^{+}_{1}=1$ and 
$\delta^{-}_{1}=0$. Using the equation (\ref{psi eq glnglm univ}) we 
derive the relation:
\begin{equation}
\Psi^{P_{\pi\bar\pi}}_{Q \Pi_{i i+1}}[\mathbf{x} + 
\mathbf{e}_{Q(i)}-\mathbf{e}_{Q(i+1)}, \overline A] -  
\Psi^{P_{\pi\bar\pi}}_{Q} [\mathbf{x} + \mathbf{e}_{Q(i)}, \overline 
A] = 0  
\end{equation}
that implies the following condition on $\Phi_{\overline A}(\hat{P} 
Q, 
\hat{P}^{-1})$:
\begin{equation}
\Phi_{\overline A}(\Pi_{\hat{P}Q(i) \hat{P}Q(i+1)}\hat{P} Q, 
\hat{P}^{-1}) = e^{i k_{\hat{P}Q(i+1)}} \Phi_{\overline A}(\hat{P} Q, 
\hat{P}^{-1})\,.
\label{int pi pibar glnglm univ}
\end{equation}

\medskip

We rewrite now all obtained conditions (\ref{WF sym glnglm 
univ}), (\ref{Phi'=-Phi glnglm univ})-(\ref{Phi=MPhi glnglm 
univ}) and (\ref{int pi pibar glnglm univ}) in a more compact form. 
Let us introduce, for $P' \equiv \hat{P}Q \in \fS_{N}$ and $Q' \equiv
\hat{P}^{-1} \in \fS_{N}$,
\begin{equation}
\hat{\Phi}(P') \equiv \sum_{Q',\overline{A}}
\Phi_{\overline{A}}(P',Q')\;\prod_{i=1}^{N} e^{A_{Q'(i)}}_{i}
\end{equation}
where the summation is over all types of excitations and all
corresponding sectors. The vector $\prod_{i=1}^{N}
e^{A_{Q'(i)}}_{i}$
belongs to $V_{1} \otimes ... \otimes V_{N}$, where 
$V$ is spanned by $\{e^{2\uparrow},e^{3\uparrow},...,e^{\fn 
\uparrow},e^{2\downarrow},e^{3\downarrow},...,e^{\fm \downarrow}\}$ 
and represents one type of $N$
excitations. The ordering of the particles is chosen such that, for 
$Q'=id$, the vector $\prod_{i=1}^{N}
e^{A_{i}}_{i}$ is taken as in (\ref{excited st glnglm}):
\begin{equation}
\prod_{i=1}^{N}
e^{A_{i}}_{i} =  \overbrace{e^{2 \uparrow} \otimes ... \otimes e^{2 
\uparrow}}^{N_{\uparrow 2}}\otimes...\otimes \overbrace{e^{\fn 
\uparrow} \otimes ... \otimes e^{\fn \uparrow}}^{N_{\uparrow \fn}} 
\otimes \overbrace{e^{2 \downarrow} \otimes ... \otimes e^{2 
\downarrow}}^{N_{\downarrow 2}} \otimes ... \otimes \overbrace{e^{\fm 
\downarrow} \otimes ... \otimes e^{\fm \downarrow}}^{N_{\downarrow 
\fm}} 
\end{equation}
To clarify the notation for $\hat{\Phi}(P')$, we give as an example 
the case $N=2$: 
\begin{eqnarray}
\hat{\Phi}(P') &=& \sum_{j=2}^\fn \Phi_{(j\uparrow,j\uparrow)}(P',id)
\,e^{j\uparrow}_{1} e^{j\uparrow}_{2} +
\sum_{j=2}^\fm \Phi_{(j\downarrow,j\downarrow)}(P',id)
\,e^{j\downarrow}_{1} e^{j\downarrow}_{2}
\nonumber \\
&&+\sum_{2\leq j<k\leq \fn}\left(
\Phi_{(j\uparrow,k\uparrow)}(P',id)
\,e^{j\uparrow}_{1} e^{k\uparrow}_{2} +
\Phi_{(j\uparrow,k\uparrow)}(P',\Pi_{12})
\,e^{k\uparrow}_{1}e^{j\uparrow}_{2}  \right)
\nonumber \\
&&+
\sum_{2\leq j<k\leq \fm}\left(
\Phi_{(j\downarrow,k\downarrow)}(P',id)
\,e^{j\downarrow}_{1}e^{k\downarrow}_{2} 
+\Phi_{(j\downarrow,k\downarrow)}(P',\Pi_{12})
\,e^{k\downarrow}_{1}e^{j\downarrow}_{2} 
\right)\nonumber \\
&&+\sum_{j=2}^\fn\sum_{k=2}^\fm \left(
\Phi_{(j\uparrow,k\downarrow)}(P',id)
\,e^{j\uparrow}_{1} e^{k\downarrow}_{2}+
\Phi_{(j\uparrow,k\downarrow)}(P',\Pi_{12})
\,e^{k\downarrow}_{1} e^{j\uparrow}_{2} 
\right)
\end{eqnarray}
Then, all the relations can be expressed in a matrix form. 
For $N=2$, it reads
\begin{equation}
\hat{\Phi}(\Pi_{12}P) = S^{(1)}_{12}(k_{1},k_{2}) 
\hat{\Phi}(P)
\end{equation}
where $S^{(1)}_{12}(k_{1},k_{2})$ (simply denoted by $S^{(1)}_{12}$ 
in the
following) acts on elementary vectors $e^{A_{1}}_{1} e^{A_{2}}_{2}$ as

1) for $\pi$-particles: 
\begin{eqnarray} 
&& S^{(1)}_{12}\; 
e^{A_1}_{1} e^{A_2}_{2} = - e^{A_2}_{1} 
e^{A_1}_{2},\;\;\text{for}\;\; A_1,A_2 \in 
\cN_{\uparrow(\downarrow)}\nonumber \\
&& S^{(1)}_{12}\;
e^{A_1}_{1} e^{A_2}_{2} =  e^{A_1}_{1} 
e^{A_2}_{2},\;\;\text{for}\;\; A_1 \in 
\cN_{\uparrow(\downarrow)},\;A_2 \in \cN_{\downarrow(\uparrow)}  
\label{Smatrix 2pi glnglm univ}
\end{eqnarray}

2) for $\bar\pi$-particles:
\begin{eqnarray}
&& S^{(1)}_{12}\;
e^{\bar A_1}_{1} e^{\bar A_2}_{2} = - e^{\bar A_2}_{1} 
e^{\bar A_1}_{2}, \;\;\text{for}\;\; \bar A_1,\bar A_2 \in \overline 
\cN_{\uparrow(\downarrow)} \nonumber \\ 
&& S^{(1)}_{12}\;
e^{\bar A_1}_{1} e^{\bar A_2}_{2} = t_{12}\,
e^{\bar A_1}_{1}e^{\bar A_2}_{2} + r_{12} \,
e^{\bar A_2}_{1}e^{\bar A_1}_{2},\;\;\text{for}\;\; \bar A_1 \in 
\overline \cN_{\uparrow(\downarrow)},\;\bar A_2 \in \overline 
\cN_{\downarrow(\uparrow)}   
\label{Smatrix 2pibar glnglm univ}
\end{eqnarray}

3) for mixed $\pi$ and $\bar\pi$-particles:
\begin{eqnarray}
&& S^{(1)}_{12}\;
e^{A_1}_{1} e^{\bar A_2}_{2} = e^{-i k_{1}}e^{A_1}_{1} 
e^{\bar A_2}_{2}\;\;\text{and}\;\; S^{(1)}_{12}\;
e^{\bar A_2}_{1}e^{A_1}_{2} = e^{i k_{2}} e^{\bar A_2}_{1} 
e^{A_1}_{2},\;\;\text{for}\;\; A_1 \in 
\cN_{\uparrow(\downarrow)},\;\bar A_2 \in \overline 
\cN_{\uparrow(\downarrow)}  \nonumber \\
&& S^{(1)}_{12}\;
e^{A_1}_{1} e^{\bar A_2}_{2} = e^{A_1}_{1} 
e^{\bar A_2}_{2}\;\;\text{and}\;\; S^{(1)}_{12}\;
e^{\bar A_2}_{1} e^{A_1}_{2} = e^{\bar A_2}_{1} 
e^{A_1}_{2},\;\;\text{for}\;\; A_1 \in 
\cN_{\uparrow(\downarrow)},\;\bar A_2 \in \overline 
\cN_{\downarrow(\uparrow)}  
\label{Smatrix pi pibar glnglm univ}
\end{eqnarray}
The parameters $t_{12}$, $r_{12}$ are defined in (\ref{r,t glnglm 
univ}) and $\cN_{\uparrow(\downarrow)}$, $\bar 
\cN_{\uparrow(\downarrow)}$ are given in (\ref{intro cN def}). \\
The notation $A \in \cN_{\uparrow(\downarrow)}$ means 
that the excitation $A$ can be equal to any value in 
$\cN_{\uparrow(\downarrow)}$ except the value $1 
\uparrow(\downarrow)$ since it was chosen as the vacuum at first 
level. 

\medskip

For an arbitrary $N$ excitations number, we write
\begin{equation}
\hat{\Phi}(\Pi_{ab}P) = S^{(1)}_{ab}(k_{a},k_{b})
\hat{\Phi}(P),
\label{Phi' =SPhi glnglm univ}
\end{equation}
where the matrix $S^{(1)}_{ab}(k_{a},k_{b})$ has the same meaning as 
above
but acts nontrivialy only on the $V_{a} \otimes V_{b}$ vector space. 
It can be written using 
permutations
and projectors:
\begin{eqnarray}
&& S^{(1)}_{ab}(k_{a},k_{b})\ =\ - \Big( P^{ \cN_{\uparrow},
\cN_{\uparrow}}_{ab} + P^{ \overline \cN_{\uparrow}, \overline
\cN_{\uparrow}} _{ab}+ P^{\cN_{\downarrow}, \cN_{\downarrow}}_{ab} + 
P^{\overline \cN_{\downarrow}, \overline \cN_{\downarrow}}_{ab} \Big) + 
e^{-ik_{a}} \Big( Id^{ \cN_{\uparrow}, \overline \cN_{\uparrow}}_{ab} +
Id^{\cN_{\downarrow}, \overline \cN_{\downarrow}}_{ab} \Big) + 
\nonumber \\ 
&&\qquad + e^{i k_{b}} \Big( Id^{ \overline \cN_{\uparrow}, 
\cN_{\uparrow}}_{ab}
+ Id^{ \overline \cN_{\downarrow} , \cN_{\downarrow}}_{ab} \Big) + 
Id^{\cN_{\uparrow} , \cN_{\downarrow} \cup \overline 
\cN_{\downarrow}}_{ab} +
Id^{ \cN_{\downarrow} , \cN_{\uparrow} \cup \overline \cN_{\uparrow }
}_{ab} + Id^{ \overline \cN_{\uparrow} , \cN_{\downarrow}}_{ab} + Id^{
\overline \cN_{\downarrow} , \cN_{\uparrow} }_{ab} + \nonumber \\ 
&&\qquad + \Big( t_{ab} Id^{ \overline \cN_{\uparrow}, \overline 
\cN_{\downarrow}
}_{ab} + r_{ab} P^{ \overline \cN_{\uparrow} , \overline 
\cN_{\downarrow}
}_{ab} \Big) + \Big( t_{ab} Id^{ \overline \cN_{\downarrow} , 
\overline
\cN_{\uparrow} }_{ab} + r_{ab} P^{ \overline \cN_{\downarrow} , 
\overline
\cN_{\uparrow} }_{ab} \Big)
\end{eqnarray}
where we defined
\begin{equation}
Id^{\cL, \cL'}_{ab} = \sum_{i \in \cL} \sum_{j \in \cL'} E_{a}^{ii} 
E_{b}^{jj} \quad \text{and} \quad 
P^{\cL, \cL'}_{ab} = \sum_{i \in \cL} 
\sum_{j \in \cL'} E_{a}^{ij} E_{b}^{ji}\,,\quad
\cL,\cL'=\cN_{\uparrow},\cN_{\downarrow},
\overline\cN_{\uparrow},\overline\cN_{\downarrow}.
\end{equation}
The matrix $S^{(1)}_{ab}(k_{a},k_{b})$ satisfies the Yang--Baxter 
equation (\ref{eq:YBE}).
% \begin{equation}
% S^{(1)}_{12}(k_{1},k_{2})
% S^{(1)}_{13}(k_{1},k_{3})
% S^{(1)}_{23}(k_{2},k_{3})
% =S^{(1)}_{23}(k_{2},k_{3})
% S^{(1)}_{13}(k_{1},k_{3})
% S^{(1)}_{12}(k_{1},k_{2})
% \label{eq:YBES}
% \end{equation}

\medskip

Now we can write the periodic boundary conditions. 
Let $C_{N}=\Pi_{N 1}...\Pi_{N N-1}$ be
a circular permutation. The sites 
$L+1$ and $1$ being identified,  we have 
 the periodicity conditions:
\begin{equation}
\Psi^{P_{\pi\bar\pi}}_{QC_{N}}[\mathbf{x}-\mathbf{e}_{Q(N)}L, 
\overline A]=\Psi^{P_{\pi\bar\pi}}_{Q}[\mathbf{x}, \overline A]
\,.
\end{equation}
In terms of $\hat{\Phi}(P)$ this yields the condition
\begin{equation}
\hat{\Phi}(PC) = e^{ik_{P(N)}L}\hat{\Phi}(P) \,.
\label{periodicity glnglm univ}
\end{equation}
If we choose $P=C_N^{N-j}$ with $j=1,\ldots,N$, we can derive a 
system of
equations on the coefficients $\hat{\Phi}(id)$ which is called 
the "auxiliary problem":
\begin{equation}
\fh_{j}\,
\hat{\Phi}(id) = e^{ik_{j}L} \hat{\Phi}(id) 
\mb{with} \fh_{j}=
S^{(1)}_{j+1,j}...S^{(1)}_{Nj}\, S^{(1)}_{1j}...S^{(1)}_{j-1,j}\,,
\qquad j=1,...,N
\label{AuxPr1 glnglm univ}
\end{equation}
where we omitted the arguments of the
$S$-matrices, $S^{(1)}_{ab} \equiv S^{(1)}_{ab}(k_{a},k_{b})$. \\
The Yang--Baxter equation for the $S$ matrix implies 
that
$[\fh_{j}\,,\,\fh_{k}]=0$ for all $j,k$, so that the new Hamiltonians
$\fh_{j}$ can be simultaneously diagonalized: we do it in the folowing
section.

\subsection{Auxiliary problem, level two}

In order to simplify the calculations, we make the change: 
$S^{(1)}_{12}\rightarrow - S^{(1)}_{12} \equiv S_{12}$. 
The eigenvalue problem to be solved reads 
\begin{equation}
S_{j+1,j}...S_{N j} S_{1 j}...S_{j-1,j}
\phi = \Lambda_{j} \phi\,.
 \label{SSSphi=Lphi glnglm univ}
\end{equation}
Again we use the coordinate Bethe ansatz. At this level we have 
$\fn+\fm-2$
types of different excitations: $e^{2\uparrow},...,e^{\fn\uparrow},
e^{2\downarrow},...,e^{\fm \downarrow}$. We choose  as the reference state 
(pseudovacuum):
\begin{equation}
\phi_{0}^{(1)} = \prod^{N}_{k=1} e^{\fm \downarrow}_{k}
\end{equation}
The eigenvectors for this auxiliary problem are given by
\begin{eqnarray}
\phi^{(1)}[\bar{A}] &=& \sum_{\mathbf{x} \in [1,N]}
\Psi^{(1)}[\mathbf{x},\bar{A}] \prod_{n=1}^{K} e^{A_{n}}_{x_{n}} 
\phi_{0}^{(1)}
\label{excited st aux pr1 glnglm univ}
\\
K&=&\sum_{i=2}^{\fp} N_{\uparrow i}+\sum_{i=2}^{\fq} 
N_{\downarrow i}+ \sum_{i=\fp+1}^{\fn} N_{\uparrow i}+ 
\sum_{i=\fq+1}^{\fm-1} N_{\downarrow i}
\,,\nonu
\displaystyle 
\bar{A} &=&
(\underbrace{2\uparrow,...,2\uparrow}_{N_{\uparrow 2}},...,
\underbrace{\fn\uparrow,...,\fn\uparrow}_{N_{\uparrow 
\fn}},\underbrace{2\downarrow,...,2\downarrow}_{N_{\downarrow 2}},...,
\underbrace{(\fm-1)\downarrow,...,(\fm-1)\downarrow}_{N_{\downarrow 
(\fm-1)}})\,.
\nonumber
\end{eqnarray}
 
The ansatz for $\Psi^{(1)}[\mathbf{x},\bar{A}] \equiv 
\Psi_{Q}[\mathbf{x},\bar{A}]$ can be written as, for 
$x_{Q(1)}<x_{Q(2)}<...<x_{Q(K)}$ and $Q \in 
\fS_{K}$, 
\begin{equation}
\Psi_{Q}[\mathbf{x},\bar{A}] = \sum_{P} \Phi_{\bar{A}}(P Q, P^{-1}) 
\prod_{i \in \cM_{\pi \uparrow}} h_{x_{i}}(a_{P(i)}) \prod_{i \in 
\cM_{\bar \pi \uparrow}} f_{x_{i}}(\lambda_{P(i)}) \prod_{i \in 
\cM_{\pi \downarrow}} \bar h_{x_{i}}(a_{P(i)}) \prod_{i \in 
\cM_{\bar \pi \downarrow}} 
g_{x_{i}}(b_{P(i)}),
\label{ansatz aux pr1 glnglm univ}
\end{equation} 
{where} $P= P_{\pi\uparrow} P_{\bar\pi\uparrow} P_{\pi\downarrow}
P_{\bar\pi\downarrow}$ is a factorized permutation such that $P_{\pi
\uparrow}$, $P_{\bar\pi\uparrow}$, $P_{\pi\downarrow}$ and $P_{\bar\pi
\downarrow}$ are the sets of permutations in $\fS_{K}$ permuting only
particles of type $\pi\uparrow$, $\bar\pi\uparrow$, $\pi\downarrow$ 
and
$\bar \pi\downarrow$ respectively, e.g. $P_{A}(i)=i$ for $i \notin 
\cM_{A}$
and $P_{A}(j)\in \cM_{A}$ for $j\in \cM_{A}$ with $A=\pi\uparrow$, 
$\bar\pi
\uparrow$, $\pi\downarrow$, $\bar\pi\downarrow$.
We recall that the sets $\cM_{\pi_{\uparrow}}$, $\cM_{\bar
\pi_{\uparrow}}$, $\cM_{\pi_{\downarrow}}$ and $\cM_{\bar
\pi_{\downarrow}}$ are defined in section \ref{sec:BAE}.
 
The eigenfunctions $f_{x}(\lambda)$, $g_{x}(b)$, $h_{x}(a)$
and $\bar h_{x}(a)$ 
correspond to one-particle solutions and are defined as
\begin{equation}
f_{x}(\lambda) = \prod^{x-1}_{m=1} \left(-\frac{i \sin{k_{m}} + i 
\lambda +
\frac{\fu}{4}}{i \sin{k_{m+1}} + i \lambda - \frac{\fu}{4}} \right)
\mb{;}
 g_{x}(b) = b^{x}
\mb{;}
\bar h_{x}(a)=h_{x}(a) = \delta(x-a).\ 
\label{fct glnglm univ} 
\end{equation}

Although the functions $h$ and $\bar h$ are identical, they will be
associated to different kinds of particles ($\pi\uparrow$ and
$\pi\downarrow$ respectively). Since they will lead to different
Hamiltonian eigenvalues (see below (\ref{Aux 1ex eigenval})), we have
chosen to distinguish them in (\ref{ansatz aux pr1 glnglm univ}). The 
form
of the eigenfunctions (\ref{fct glnglm univ}) 
is supported by the fact that the eigenfunction in (\ref{AuxPr1 glnglm
univ}) should be independent of the index $j$.

We can remark that the Bethe roots $a_{i}$ for $i \in 
\cM_{\pi_{\downarrow}}$ or $\cM_{\pi_{\uparrow}}$ corresponding to 
$\pi_{\downarrow}$ and $\pi_{\uparrow}$ particles are 
already quantized on the small chain. In order to have an independent 
number of eigenvectors (\ref{excited st aux pr1 glnglm univ}), one 
should take the following conditions into account:
\begin{equation}
a_{i}< a_{i+1} \;\;\text{for}\;\; i \in 
\cM_{\pi_{\uparrow(\downarrow)}}
\end{equation}
that leads to the fact that in (\ref{ansatz aux pr1 glnglm 
univ}),
$P_{\pi \uparrow}$ and $P_{\pi \downarrow}$ are fixed: 
$P_{\pi_{\uparrow(\downarrow)}} Q (j) = \text{const}$ (see below).

The action of the auxiliary Hamiltonian (\ref{AuxPr1 glnglm univ}) on 
the wavefunction can be calculated using the relations given in the
previous section. In the recursive representation of the 
Hamiltonian (for details, see \cite{sutherland})
\begin{equation}
S_{j-k,j} \ldots S_{j-1,j} 
\phi^{(1)}[\bar{A}] \equiv \sum_{\mathbf{x} \in [1,N]}
\hat H^{(k)}_{j} \Psi_{Q}[\mathbf{x},\bar{A}] \prod_{n=1}^{K} 
e^{A_{n}}_{x_{n}} \phi_{0}^{(1)} 
\label{recursHam glnglm univ}
\end{equation} 
one can derive the following recursive relations between the 
coefficients
$\hat H^{(k)}_{j} \Psi_{Q}[\mathbf{x},\bar{A}]$ for decreasing $k$. 
Nontrivial relations occur when $\hat H^{(k)}_{j}$ acts on 
$\Psi_{Q}[\mathbf{x},\bar{A}]$ for which some coordinate $x_p$ is 
equal to 
$j$ and/or $j-k$. One gets two sets of relations. The first one reads:
\begin{eqnarray*}
\hat H^{(k)}_{j} \Psi_{Q}[\mathbf{x'},\ovw{x_{p}}{j};\bar{A}] = 
\begin{cases} 
-t_{j-k,j} \hat H^{(k-1)}_{j} 
\Psi_{Q}[\mathbf{x'},\ovw{x_{p}}{j};\bar{A}] - r_{j-k,j} \hat 
H^{(k-1)}_{j} \Psi_{Q'}[\mathbf{x'},\ovw{x_{p}}{j-k};\bar{A}] 
&\text{for}\;\;p \in \cM_{\bar \pi \uparrow} \\
\hat H^{(k-1)}_{j} \Psi_{Q'}[\mathbf{x'},\ovw{x_{p}}{j-k};\bar{A}] 
&\text{for}\;\;p \in \cM_{\bar \pi \downarrow} \\
-\hat H^{(k-1)}_{j} \Psi_{Q}[\mathbf{x'},\ovw{x_{p}}{j};\bar{A}] 
&\text{for}\;\;p \in \cM_{\pi \uparrow} \\
-e^{i k_{j}}\hat H^{(k-1)}_{j} 
\Psi_{Q}[\mathbf{x'},\ovw{x_{p}}{j};\bar{A}] &\text{for}\;\;p \in 
\cM_{\pi \downarrow} 
\end{cases}
\end{eqnarray*}
\begin{equation}  
\null  \label{eq:relrec1}
\end{equation}
The above relations are invariant if we exchange $j$ with $j-k$ except 
the last one, that is transformed as follows:
\begin{equation}
\hat H^{(k)}_{j} \Psi_{Q}[\mathbf{x'},\ovw{x_{p}}{j-k};\bar{A}] = 
-e^{-i k_{j-k}}\hat H^{(k-1)}_{j} 
\Psi_{Q}[\mathbf{x'},\ovw{x_{p}}{j-k};\bar{A}] \mb{for}p 
\in \cM_{\pi \downarrow}
\label{eq:relrec2}
\end{equation}
The second set of relations is
\begin{eqnarray*}
\hat H^{(k)}_{j} 
\Psi_{Q}[\mathbf{x'},\ovw{x_{p_{1}}}{j},\ovw{x_{p_{2}}}{j-k};\bar{A}] 
= \begin{cases} 
\hat H^{(k-1)}_{j} \Psi_{Q'}[\mathbf{x'},j-k,j;\bar{A}],\;\;\; 
\text{for}\;\; A_{p_{1}},A_{p_{2}} \in \cM_{\pi \sigma}, \cM_{\bar 
\pi \sigma}\\ 
-t_{j-k,j} \hat H^{(k-1)}_{j} \Psi_{Q}[\mathbf{x'},j,j-k;\bar{A}] - 
r_{j-k,j} \hat H^{(k-1)}_{j} \Psi_{Q'}[\mathbf{x'},j-k,j;\bar{A}], \\ 
\qquad \qquad \qquad \qquad \qquad \qquad \qquad \text{for}\;\;p_{1} 
\in \cM_{\bar \pi \uparrow(\downarrow)},\;p_{2} \in \cM_{\bar 
\pi \downarrow(\uparrow)} \\ 
-e^{-i k_{j-k}} \hat H^{(k-1)}_{j} 
\Psi_{Q}[\mathbf{x'},j,j-k;\bar{A}],\;\;\;\text{for}\;\;p_{1} \in 
\cM_{\bar \pi \uparrow(\downarrow)},\;p_{2} \in \cM_{ \pi 
\uparrow(\downarrow)} \\
-e^{i k_{j}} \hat H^{(k-1)}_{j} 
\Psi_{Q}[\mathbf{x'},j,j-k;\bar{A}],\;\;\;\text{for}\;\;p_{1} \in 
\cM_{ \pi \uparrow(\downarrow)},\;p_{2} \in \cM_{\bar \pi 
\uparrow(\downarrow)} \\
- \hat H^{(k-1)}_{j} 
\Psi_{Q}[\mathbf{x'},j,j-k;\bar{A}],\;\;\;\text{for}\;\;p_{1} \in 
\cM_{\bar \pi \uparrow(\downarrow)},\;p_{2} \in \cM_{ \pi 
\downarrow(\uparrow)} \\ 
- \hat H^{(k-1)}_{j} 
\Psi_{Q}[\mathbf{x'},j,j-k;\bar{A}],\;\;\;\text{for}\;\;p_{1} \in 
\cM_{\pi \uparrow(\downarrow)},\;p_{2} \in \cM_{\bar \pi 
\downarrow(\uparrow)} \\ 
- \hat H^{(k-1)}_{j} 
\Psi_{Q}[\mathbf{x'},j,j-k;\bar{A}],\;\;\;\text{for}\;\;p_{1} \in 
\cM_{\pi \downarrow(\uparrow)},\;p_{2} \in \cM_{ \pi 
\uparrow(\downarrow)} 
\end{cases}
\end{eqnarray*}
\begin{equation}  
  \label{eq:relrec3}
\end{equation}
In eqs. (\ref{eq:relrec1})--(\ref{eq:relrec3}), the vector 
$\mathbf{x'}$ 
corresponds to the vector $\mathbf{x}$ without the components $x_p$ 
or 
$x_{p_1},x_{p_2}$ and does not contain any position $x_q = j$ or 
$j-k$. \\
Using these relations, we can then apply the whole product of 
$S$-matrices
(\ref{SSSphi=Lphi glnglm univ}) on a one-excitation function. We get 
for
$x\neq j$
\begin{eqnarray}
&& \hat H^{(N-1)}_{j} f_{x}(\lambda) = \sigma_{j}(\lambda) 
f_{x}(\lambda)
\quad \text{and} \quad \hat H^{(N-1)}_{j}g_{x}(b) = b g_{x}(b) 
\nonumber \\ 
&& \hat H^{(N-1)}_{j}h_{x}(a) = - h_{x}(a) \quad \text{and} \quad \hat
H^{(N-1)}_{j} \bar h_{x}(a) =-e^{- i k_{x}} \bar h_{x}(a)
\label{Aux 1ex eigenval}
\end{eqnarray}
and for $x=j$
\begin{eqnarray}
&& \hat H^{(N-1)}_{j} f_{j}(\lambda) = \prod_{l=1,l\neq j}^{N}
\sigma_{l}(\lambda) f_{j}(\lambda) \quad \text{and} \quad \hat 
H^{(N-1)}_{j}
g_{j}(b) = b^{1-N} g_{j}(b) \nonumber \\
&& \hat H^{(N-1)}_{j} h_{j}(a) =(-1)^{N-1} h_{j}(a) \quad \text{and} 
\quad
\hat H^{(N-1)}_{j} \bar h_{j}(a) = (-1)^{N-1} e^{i k_{j}(N-1)} \bar
h_{j}(a)\quad
\end{eqnarray}
where 
\begin{equation}
\sigma_{j}(\lambda)= - \frac{i \sin(k_{j}) + i \lambda + 
\frac{\fu}{4}}{i
\sin(k_{j}) + i\lambda - \frac{\fu}{4}}
\end{equation}

Now we will consider the $K$-excitations eigenvector (\ref{excited st 
aux
pr1 glnglm univ}) with the ansatz (\ref{ansatz aux pr1 glnglm univ}). 
There
are three different possible cases:
\begin{itemize}
\item[I)] there exists a Bethe 
root $a_{l} = j$ for some $l \in \cM_{\pi \uparrow}$, with $j$ 
being the 
index in (\ref{AuxPr1 glnglm univ}), 
\item[II)] there exists another Bethe 
root $a_{m} = j$ for some $m \in \cM_{\pi \downarrow}$, 
\item[III)] there is no such Bethe roots. 
\end{itemize}
We detail the calculations for each of these cases.

\subsubsection*{I) There exists a Bethe root $a_{l} = j$ for some $l 
\in
\cM_{\pi \uparrow}$} 

In this first case, let us introduce a set of integers $\{ \alpha_{i} 
\in
[1,K]\}_{i \in \cM_{\pi \uparrow}}$ such that we have $Q(\alpha_{i}) 
\in
\cM_{\pi \uparrow}$. The set is ordered: $\alpha_{i} < \alpha_{i+1}$. 
We
also define $\alpha_{l}$ such that $x_{Q(\alpha_{l})} = j$.

The Hamiltonian acting on the wave function gives the following 
result:
\begin{eqnarray}
&&\hat H^{(N-1)}_{j} \Psi_{Q}(\mathbf x; 
x_{Q(\alpha_{1})},...,x_{Q(\alpha_{l-1})},
\ovw{x_{Q(\alpha_{l})}}{j},x_{Q(\alpha_{l+1})},...,
x_{Q(\alpha_{N_{\uparrow \pi}})}) 
\nonumber \\
&&\qquad = (-1)^{N-1-N_{\uparrow \pi}} e^{i k_{j} N_{\uparrow \bar 
\pi}} \,
\Psi_{Q C_{\pi \uparrow}}(\mathbf x; 
x_{Q(\alpha_{2})},...,\ovw{x_{Q(\alpha_{l})}}{j},x_{Q(\alpha_{l+1})},
x_{Q(\alpha_{l+2})},...,x_{Q(\alpha_{1})}) 
 \nonumber \\
&&\qquad =
\Lambda_{j} \Psi_{Q}(\mathbf x; 
x_{Q(\alpha_{1})},...,x_{Q(\alpha_{l-1})},j,x_{Q(\alpha_{l+1})},...,
x_{Q(\alpha_{N_{\uparrow \pi}})}) 
\end{eqnarray}
and finally we get the condition on $\Phi_{\bar{A}}(P Q, P^{-1})$:
\begin{eqnarray}
\frac{\Phi_{\bar{A}}(P' Q C_{\pi \uparrow}, P'^{-1}) 
}{\Phi_{\bar{A}}(P Q, P^{-1})} = \Lambda_{j} (-1)^{N-1-N_{\uparrow 
\pi}} e^{-i k_{j} N_{\uparrow \bar \pi}}\,.
\label{aux period pi up}
\end{eqnarray} 
In equality (\ref{aux period pi up}), we have introduced $C_{\pi 
\uparrow} = 
\Pi^{\alpha_{1}}_{\alpha_{N_{\uparrow 
\pi}}}...\Pi^{\alpha_{N_{\uparrow 
\pi} - 1 }}_{\alpha_{N_{\uparrow \pi}}}$ and for
 $P=P_{\pi \uparrow}P_{\pi \downarrow}P_{\bar \pi 
\uparrow}P_{\bar \pi \downarrow}$, $P'=P'_{\pi \uparrow}P_{\pi 
\downarrow}P_{\bar \pi \uparrow}P_{\bar \pi \downarrow}$ where 
$P'_{\pi \uparrow} = \Pi^{N_{\uparrow 
\pi} - 1 }_{N_{\uparrow 
\pi}} ...\Pi^{1}_{N_{\uparrow 
\pi}} P_{\pi \uparrow} $.

% The above condition $P'_{\pi \uparrow} = (...) P_{\pi \uparrow} $ can 
% be obtained if we use the fact 
% already mentioned above. For ordered integers $a_{i}< a_{i+1} 
% \;\;\text{for}\;\; i \in \cM_{\pi_{\uparrow}}$ and ordered 
% coordinates $x_{Q(i)}<x_{Q(i+1)}$ for any $i$ in (\ref{ansatz aux pr1 
% glnglm univ}) we have $P_{\pi \uparrow}Q(\alpha_{i})=i$ for $i 
% \in \cM_{\pi \uparrow}$ and any sector $Q$ (and $P_{\pi \uparrow}$ 
% depending on $Q$). Therefore we can get 
% the relation for two sectors $Q$ and $Q C_{\pi \uparrow}$.
% 
% The only condition that relates the coefficients $\Phi_{\bar{A}}(P Q,
% P^{-1})$ is the one obtained for identical particles. 
If we change the
sector $Q \rightarrow Q \Pi_{ab}$, which means that we interchange two
identical particles $x_{Q(a)}$ and $x_{Q(b)}$, the coefficient
$\Phi_{\bar{A}}(P Q \Pi_{ab}, P^{-1})$ should remain invariant:
\begin{equation}
\Phi_{\bar{A}}(P' Q \Pi_{ab}, P'^{-1}) = \Phi_{\bar{A}}(P Q, 
P^{-1}) \quad \text{with} \quad Q(a),Q(b)\in \cM_{\pi \uparrow}
\end{equation}
where $P'=P'_{\pi \uparrow}P_{\pi \downarrow}P_{\bar \pi 
\uparrow}P_{\bar \pi \downarrow}$ and $P'_{\pi \uparrow}$ is such 
that $P'_{\pi \uparrow} = P_{\pi \uparrow}\Pi_{Q(a)Q(b)}$. \\
The relation (\ref{aux period pi up}) can be simplified to:
\begin{eqnarray}
\frac{\Phi_{\bar{A}}(P Q , P^{-1} \tilde{C}_{\pi \uparrow}) 
}{\Phi_{\bar{A}}(P Q, P^{-1})} = \Lambda_{j} (-1)^{N-N_{\uparrow 
\pi}} e^{-i k_{j} N_{\uparrow \bar \pi}}
\label{aux period pi up 2}
\end{eqnarray} 
with $\tilde{C}_{\pi \uparrow}=\Pi^{1}_{N_{\uparrow \pi}} ... 
\Pi^{N_{\uparrow \pi} - 1 }_{N_{\uparrow \pi}}$. 
Now, taking the product of this equation by itself $N_{\uparrow \pi}$
times and changing $P \rightarrow \tilde{C}_{\pi \uparrow} P 
\rightarrow
\tilde{C}^{2}_{\pi \uparrow} P\to\ldots\to \tilde{C}_{\pi
\uparrow}^{N_{\uparrow \pi}}P$, we obtain
\begin{eqnarray}
\Lambda_{j} (-1)^{N-N_{\uparrow \pi}} e^{-i k_{j} N_{\uparrow \bar 
\pi}} =
e^{2 \pi i \frac{m_{ \uparrow \pi}}{N_{\uparrow \pi}}} \qquad
(m_{ \uparrow \pi}=1,...,N_{\uparrow \pi})
\end{eqnarray} 
It determines the eigenvalue of the auxiliary problem for spin up
$\pi$-particles. We also recall that the result has to be multiplied by
$(-1)^{N-1}$ in order to come back to the initial problem due to the 
sign
change done at the beginning of the section, see (\ref{SSSphi=Lphi 
glnglm
univ}).
% \\
% For instance we have eigenvectors $\phi^{(1)}[\bar{A}]$ with 
% multiplicity
% in the $\pi\uparrow$ sector equal to $\frac{N!}{(N-K)! (N_{\uparrow
% \pi})!}$ and constant coefficients $\Phi_{\bar{A}}(P_{\pi 
% \uparrow}\hat P
% Q, (P_{\pi \uparrow}\hat P)^{-1})$ for any $P_{\pi \uparrow}$ and 
% $\hat P$
% fixed. If we take all coefficients $\Phi_{\bar{A}}(P_{\pi 
% \uparrow}\hat P
% Q, (P_{\pi \uparrow}\hat P)^{-1}) = \Phi_{\bar{A}}(\hat P Q, \hat 
% P^{-1})$
% (we noted $\hat P =P_{\pi \downarrow}P_{\bar \pi \uparrow}P_{\bar \pi
% \downarrow} $), equation (\ref{aux period pi up}) shows that the 
% eigenvalue
% $\Lambda_{j}$ is equal to
% \begin{equation}
% \Lambda_{j} = (-1)^{N-N_{\uparrow \pi}} e^{i k_{j} N_{\uparrow \bar 
% \pi}}
% \end{equation}

\subsubsection*{II) There exists another Bethe root $a_{m} = j$ for 
some $m
\in \cM_{\pi \downarrow}$} 

Following the same steps as in the first case, we obtain 
% We define a set of integer $\{ \alpha_{i} \in [1,K]\}_{i \in \cM_{\pi
% \downarrow}}$ such that we have $Q(\alpha_{i}) \in \cM_{\pi 
% \downarrow}$
% and $\alpha_{i} < \alpha_{i+1}$. We also define $\alpha_{l}$ such that
% $x_{Q(\alpha_{l})} = j$. Therefore, we have
% \begin{eqnarray}
% \frac{\Phi_{\bar{A}}(P' Q C_{\pi \downarrow}, P'^{-1}) 
% }{\Phi_{\bar{A}}(P Q, P^{-1})} = \Lambda_{j} 
% (-1)^{N-N_{\downarrow \pi}} e^{-i k_{j} (N_{\downarrow \bar 
% \pi}+N_{\downarrow \fm})}
% \end{eqnarray} 
% where $P=P_{\pi \uparrow}P_{\pi \downarrow}P_{\bar \pi 
% \uparrow}P_{\bar \pi
% \downarrow}$, $P'=P_{\pi \uparrow}P'_{\pi \downarrow}P_{\bar \pi
% \uparrow}P_{\bar \pi \downarrow}$ and $C_{\pi \downarrow} =
% \Pi^{\alpha_{1}}_{\alpha_{N_{\downarrow 
% \pi}}}...\Pi^{\alpha_{N_{\downarrow
% \pi} - 1 }}_{\alpha_{N_{\downarrow \pi}}}$ with $P'_{\pi \downarrow} =
% \Pi^{N_{\downarrow \pi} - 1 }_{N_{\downarrow \pi}}
% ....\Pi^{1}_{N_{\downarrow \pi}} P_{\pi \downarrow}$. \\
% As previously, this relation can be simplified, and we get
\begin{eqnarray}
\frac{\Phi_{\bar{A}}(P Q , P^{-1} \tilde{C}_{\pi \downarrow})}
{\Phi_{\bar{A}}(P Q, P^{-1})} = \Lambda_{j} (-1)^{N-N_{\downarrow 
\pi}}
e^{-i k_{j} (N_{\downarrow \bar \pi}+N_{\downarrow \fm})}
\label{aux period pi down 2}
\end{eqnarray} 
where $P=P_{\pi \uparrow}P_{\pi \downarrow}
P_{\bar \pi \uparrow}P_{\bar \pi\downarrow}$, 
$P'=P_{\pi \uparrow}P'_{\pi \downarrow}P_{\bar \pi\uparrow}
P_{\bar \pi \downarrow}$ and $\tilde{C}_{\pi \downarrow}=\Pi^{1}_{N_{\downarrow \pi}} ... 
\Pi^{N_{\downarrow \pi} - 1 }_{N_{\downarrow \pi}}$. \\
Finally using the same trick for the product of equations, we obtain
\begin{eqnarray}
\Lambda_{j} (-1)^{N-N_{\downarrow \pi}} e^{-i k_{j} (N_{\downarrow 
\bar
\pi}+N_{\downarrow \fm})} = e^{2 \pi i \frac{m_{\downarrow
\pi}}{N_{\downarrow \pi}}},\;\;m_{\downarrow \pi}=1,...,N_{\downarrow 
\pi}
\end{eqnarray} 

It allows to determine the eigenvalue of the auxiliary problem for 
$\pi\downarrow$ particles. Again, this result should  be 
multiplied by a factor $(-1)^{N-1}$ to come back to the 
initial problem.

\medskip

For identical particles we have the relation:
\begin{equation}
\Phi_{\bar{A}}(P' Q \Pi_{ab}, P'^{-1}) = \Phi_{\bar{A}}(P Q, 
P^{-1}),\;\;\text{with}\;\;Q(a),Q(b)\in \cM_{\pi \uparrow}
\end{equation}
where $P'=P_{\pi \uparrow}P'_{\pi \downarrow}P_{\bar \pi 
\uparrow}P_{\bar \pi \downarrow}$ and $P'_{\pi \downarrow}$ is such 
that $P'_{\pi \downarrow} = P_{\pi \downarrow}\Pi_{Q(a)Q(b)} $. 

% If we take, for example, all coefficients $\Phi_{\bar{A}}(P_{\pi
% \downarrow}\hat P Q, (P_{\pi \downarrow}\hat P)^{-1}) = 
% \Phi_{\bar{A}}(\hat
% P Q, \hat P^{-1})$ (we noted $\hat P =P_{\pi \uparrow}P_{\bar \pi
% \uparrow}P_{\bar \pi \downarrow} $), then again (\ref{excited st aux 
% pr1
% glnglm univ}) is an eigenvector with eigenvalue $\Lambda_{j}$ equal to
% \begin{equation}
% \Lambda_{j} = (-1)^{N-N_{\downarrow \pi}} e^{i k_{j} 
% (N_{\downarrow \bar \pi}+N_{\downarrow \fm})}
% \end{equation}

\subsubsection*{III) There is no such Bethe root} 

In this last case, we have $\pi$-particles with spin up or down 
distributed
anywhere on the chain but the site $j$. Acting with the Hamiltonian
(\ref{AuxPr1 glnglm univ}) on the wavefunction (\ref{ansatz aux pr1 
glnglm
univ}) we get 
\begin{eqnarray}
\hat H^{(N-1)}_{j} \Psi_{Q}[\mathbf{x},\bar{A}] = (-1)^{N_{\pi 
\uparrow}+ N_{\pi \downarrow}} \prod_{l \in \cM_{\pi \downarrow}} 
e^{- i k_{a_{l}}}  \prod_{i \in \cM_{\bar \pi \uparrow}} 
\sigma_{j}(\lambda_{i}) \prod_{i \in \cM_{\bar \pi \downarrow}} 
b_{i} \, \Psi_{Q}[\mathbf{x},\bar{A}],\;\; \text{if}\;\;\mathbf{x} 
\neq j
\label{Aux eigenval eq}
\end{eqnarray}
with the following conditions on the coefficients $\Phi_{\bar{A}}(P
Q,P^{-1})$:

1) for all $PQ(i)$,$PQ(i+1)$ in $\cM_{\bar \pi \uparrow}$ 
\begin{equation}
\frac{\Phi_{\bar{A}}(P Q \Pi_{i i+1},(P \Pi_{Q(i) Q(i+1)})^{-1})}
{\Phi_{\bar{A}}(P Q, P^{-1})} = \frac{ i\lambda_{PQ(i)} -
i\lambda_{PQ(i+1)} - \frac{\fu}{2}}{i\lambda_{PQ(i)} - 
i\lambda_{PQ(i+1)} +
\frac{\fu}{2}} \equiv \alpha^{-1}_{PQ(i) PQ(i+1)}
\label{PhiPab=Phi f f}
\end{equation}

2) for all $PQ(i)$,$PQ(i+1)$ in $ \cM_{\bar \pi \downarrow}$ we 
impose 
\begin{equation}
\frac{\Phi_{\bar{A}}(P Q \Pi_{i i+1},(P \Pi_{Q(i) Q(i+1)})^{-1} )}{ 
\Phi_{\bar{A}}(PQ, P^{-1})} = 1
\label{PhiPab=Phi g g}
\end{equation}

3) for all $PQ(i) \in \cM_{\bar \pi \downarrow}$ and  $PQ(i+1) 
\in \cM_{\bar \pi \uparrow}$
\begin{equation}
\frac{\Phi_{\bar{A}}(P Q \Pi_{i i+1}, P^{-1})}{\Phi_{\bar{A}}(P Q, 
P^{-1})} = b_{PQ(i)}^{-1}
\label{PhiPab=Phi f g}
\end{equation}
the same relation holds for all $PQ(i) \in \cM_{\bar \pi \uparrow}$, 
$PQ(i+1) \in \cM_{\bar \pi \downarrow}$ and changing $Q \rightarrow Q
\Pi_{i i+1}$.

4) for all $PQ(i) \in \cM_{\pi \uparrow}$ and $PQ(i+1) \in 
\cM_{\bar \pi \uparrow}$
\begin{equation}
\frac{ \Phi_{\bar{A}}(P Q \Pi_{i i+1}, P^{-1})} 
{\Phi_{\bar{A}}(PQ,P^{-1})}
= e^{-i k_{a_{PQ(i)}}} \sigma_{a_{PQ(i)}}(\lambda_{PQ(i+1)})
\label{PhiPab=Phi pi up pibar up}
\end{equation}

5) for all $PQ(i) \in \cM_{\pi \downarrow}$ and $PQ(i+1) \in 
\cM_{\bar \pi \uparrow}$
\begin{equation}
\frac{\Phi_{\bar{A}}(P Q \Pi_{i i+1}, P^{-1})} 
{\Phi_{\bar{A}}(PQ,P^{-1})}
= e^{i k_{a_{PQ(i)}}} \sigma_{a_{PQ(i)}}(\lambda_{PQ(i+1)})
\label{PhiPab=Phi pi down pibar up}
\end{equation}

6) for all $PQ(i) \in \cM_{\pi \uparrow}$ and $PQ(i+1) \in 
\cM_{\bar \pi \downarrow}$
\begin{equation}
\frac{\Phi_{\bar{A}}(P Q \Pi_{i i+1}, 
P^{-1})}{\Phi_{\bar{A}}(PQ,P^{-1})}
= b_{PQ(i+1)}
\label{PhiPab=Phi pi up pibar down}
\end{equation}

7) for all $PQ(i) \in \cM_{\pi \downarrow}$ and $PQ(i+1) \in 
\cM_{\bar \pi \downarrow}$
\begin{equation}
\frac{\Phi_{\bar{A}}(P Q \Pi_{i i+1} , P^{-1})} 
{\Phi_{\bar{A}}(PQ,P^{-1})}
= b_{PQ(i+1)}
\label{PhiPab=Phi pi down pibar down}
\end{equation}

In addition, for all $PQ(i)$,$PQ(i+1)$ in $\cM_{\bar \pi \uparrow}$ 
or $
\cM_{\bar \pi \downarrow}$ we impose the condition for identical 
particles:
\begin{equation}
\frac{\Phi_{\bar{A}}(P Q ,(P \Pi_{Q(i) Q(i+1)})^{-1} )}
{\Phi_{\bar{A}}(PQ,P^{-1})} = 1
\label{Aux Phi identpart}
\end{equation}

For the perodicity conditions, 
there are again two different subcases: 
\begin{itemize}
\item[III-A)] a $\bar \pi \downarrow$-particle is on the $j$ site, 
\item[III-B)] a $\bar \pi \uparrow$-particle is on the $j$ site. 
\end{itemize}
Each case leading to slightly different conditions, we treat 
them
separately.

\paragraph{III-A) A $\bar \pi \downarrow$-particle is on the $j$ 
site.} 
The periodic boundary condition on the coefficients $\Phi_{\bar{A}}(P 
Q ,P^{-1})$ comes from the relations
\begin{equation}
\hat H^{(N-1)}_{j} \Psi_{Q}[\mathbf{x},\ovw{\bar \pi 
\downarrow}{j};\bar{A}] = (-1)^{N_{\pi \uparrow}+ N_{\pi \downarrow}} 
\prod_{l \in \cM_{\pi \downarrow}} 
e^{- i k_{a_{l}}}  \prod_{i \in \cM_{\bar \pi \uparrow}} 
\sigma_{j}(\lambda_{i}) \prod_{i \in \cM_{\bar \pi \downarrow}} 
b_{i} \, \Psi_{Q}[\mathbf{x},\ovw{\bar \pi 
\downarrow}{j};\bar{A}],\;\; \text{and}\;\;\mathbf{x} \neq j
\nonumber \\
\end{equation}
Using relations (\ref{PhiPab=Phi f f}) to (\ref{Aux Phi 
identpart})
we can calculate $\hat H^{(N-1)}_{j} \Psi_{Q}[\mathbf{x},j;\bar{A}]$ 
to get the condition
\begin{equation}
\frac{\Phi_{\bar{A}}(P Q C_{K}, P^{-1})}{\Phi_{\bar{A}}(PQ,P^{-1})} =
[b_{PQ(K)}]^{N}
\label{Aux eq:period-glnglm g univ}
\mb{with}C_{K} = \Pi^{1}_ {K}...\Pi^{K-1}_{K}\,.
\end{equation}
We start with 
$Q=Q_{\bar\pi\downarrow}$
such that
\begin{equation}
Q_{\bar \pi \downarrow}(i) = 
\begin{cases} 
  i,&\text{for}\;\;i\notin \cM_{\bar \pi \downarrow}\\
  j \in \cM_{\bar \pi \downarrow},& \text{for}\;\;i\in \cM_{g}
\end{cases}
\;\;\; \Rightarrow\;\;\;
PQ(i) = 
\begin{cases} 
  m \equiv \hat P (i) \in [1,K] \cap \cM_{\bar \pi \downarrow},&\;i 
\notin
  \cM_{\bar \pi \downarrow} \\
  P_{\bar \pi \downarrow} Q_{\bar \pi \downarrow}(i) \equiv 
\tilde{P}(i)\in
  \cM_{\bar \pi \downarrow},&\;i\in \cM_{\bar
  \pi \downarrow}
\end{cases}
\end{equation}
and
\begin{equation}
PQC_{K}= \Pi^{\hat P(1)}_{\tilde{P}(K)}\Pi^{\hat 
P(2)}_{\tilde{P}(K)}...\Pi^{\hat P(K-N_{\downarrow \bar 
\pi})}_{\tilde{P}(K)}\Pi^{\tilde{P}(K-N_{\downarrow \bar \pi}+ 
1)}_{\tilde{P}(K)}...\Pi^{\tilde{P}(K-1)}_{\tilde{P}(K)}PQ\,.
\end{equation}
Using relations (\ref{PhiPab=Phi pi up pibar down}), 
(\ref{PhiPab=Phi
pi down pibar down}) and "conjugated" (\ref{PhiPab=Phi f g}) when $Q
\rightarrow Q \Pi_{i i+1}$, the coefficient
$\Phi_{\bar{A}}(P Q C_{K},P^{-1})$ simplifies to:
\begin{eqnarray}
\Phi_{\bar{A}}(P Q C_{K},P^{-1}) &\!\!=\!\!& \Phi_{\bar{A}}( 
\Pi^{\hat 
P(1)}_{\tilde{P}(K)} \Pi^{\hat P(2)}_{\tilde{P}(K)}...\Pi^{\hat P(K - 
N_{\downarrow \bar \pi})}_{\tilde{P}(K)} 
\Pi^{\tilde{P}(K - N_{\downarrow \bar \pi}+1)}_{\tilde{P}(K)}
....\Pi^{\tilde{P}(K-1)}_{\tilde{P}(K)}PQ , P^{-1})
\nonumber \\ 
&\!\!=\!\!& [b_{\tilde{P}(K)}]^{N_{\uparrow \pi} + N_{\uparrow \bar 
\pi } +
N_{\downarrow \pi} } \Phi_{\bar{A}}(\Pi^{\tilde{P}(K - N_{\downarrow 
\bar
\pi}+1)}_{\tilde{P}(K)}...\Pi^{\tilde{P}(K-1)}_{\tilde{P}(K)}PQ , 
P^{-1})\,.
\end{eqnarray}
Finally we have the equation
\begin{equation}
\Phi_{\bar{A}}( \Pi^{\tilde{P}(K - N_{\downarrow \bar 
\pi}+1)}_{\tilde{P}(K)}...\Pi^{\tilde{P}(K-1)}_{\tilde{P}(K)}PQ , 
P^{-1}) 
= [b_{\tilde{P}(K)}]^{N-(N_{\uparrow \pi} + N_{\uparrow \bar \pi } 
+ N_{\downarrow \pi})} \Phi_{\bar{A}}(PQ , P^{-1})
\end{equation}
In the next step of the procedure, we introduce the vector 
$\hat{\Phi}_{\bar \pi \downarrow}(P)$:
\begin{equation}
\hat{\Phi}_{\bar \pi \downarrow}(P) \equiv \sum_{\bar{A} \in G} \;\; 
\sum_{Q'\in \fS_{N_{\downarrow \bar \pi}}} \Phi_{\bar{A}}(P,Q') 
\prod_{i \in \cM_{\bar \pi \downarrow}} e^{\bar{A}_{Q'(i)}}_{i}
\end{equation}
where $G=\spn\{(\fq +1) \downarrow, ..., (\fm-1) \downarrow \}$. 
% The summation on $\bar{A}\in G$ is done on $\bar{A}_{i}$ 
% (defined in
% (\ref{excited st aux pr1 glnglm univ})) for $i \in \cM_{\bar \pi
% \downarrow}$ equal to all possible values in $G$. 
The particle order is
chosen such that for $Q'=id$, the vector $\prod_{i \in \cM_{\bar 
\pi
\downarrow}} e^{\bar{A}_{i}}_{i}$ is equal to $\overbrace{e^{(\fq+1)
\downarrow}\otimes ...\otimes e^{(\fq+1) \downarrow}}^{N_{\downarrow 
(\fq+1)
}}\otimes ... \otimes \overbrace{e^{(\fm-1) \downarrow}\otimes 
...\otimes
e^{(\fm-1) \downarrow}}^{N_{\downarrow (\fm-1) }}$. The permutation $Q'$ 
is the image of $Q_{\bar \pi \downarrow}$ when one shifts by
$K - N_{\downarrow \bar \pi}$ the space of integers
on which the permutation group $\fS_{N_{\downarrow \bar \pi}}$ acts.

The relations (\ref{PhiPab=Phi g g}) and (\ref{Aux Phi identpart})
can be gathered in $\hat{\Phi}_{\bar \pi \downarrow}(P)$:
\begin{equation}
\hat{\Phi}_{\bar \pi \downarrow}(\Pi_{ab}P)= \cP_{ab} 
\hat{\Phi}_{\bar \pi \downarrow}(P)
\end{equation}
where $\cP_{ab}$ is the permutation acting on the particles located 
at the $a$ and $b$ positions in $\hat{\Phi}_{\bar \pi 
\downarrow}(P)$. 
The periodicity problem can then be rewritten in the following form
\begin{equation}
\hat{\Phi}_{\bar \pi \downarrow}\Big(\Pi^{P(K - N_{\downarrow \bar 
\pi}+1)}_{P(K)}...\Pi^{P(K-1)}_{P(K)} P\Big) = [b_{P(K)}]^{N- K + 
N_{\downarrow \bar \pi}} \hat{\Phi}_{\bar \pi \downarrow}(P)
\end{equation}
which can be simplified to 
\begin{equation}
\cP_{P(K - N_{\downarrow \bar \pi}+1) P(K)}...\cP_{P(K-1) P(K)}  
\hat{\Phi}_{\bar \pi \downarrow}(P) = [b_{P(K)}]^{N-K + 
N_{\downarrow \bar \pi}} \hat{\Phi}_{\bar \pi \downarrow}(P)
\end{equation}
Choosing $P = C^{-m}_{\bar \pi \downarrow}$ with $C_{\bar 
\pi \downarrow}=\Pi^{K - N_{\downarrow \bar 
\pi}+1}_{K}...\Pi^{K-1}_{K}$, we get the following Bethe 
equation
\begin{eqnarray}
\big(b_{m}\big)^{N-K+N_{\downarrow \bar \pi}} &=& \exp\left(\frac{2 
\pi 
i}{N_{\downarrow \bar \pi}} \sum_{j=1}^{N_{\downarrow \bar \pi} - 
N_{\downarrow (\fm-1)}} n_{j}\right)\,,\ m \in \cM_{\bar \pi 
\downarrow}
\\
\mbox{with}
&&
1 \leq n_{1} < ... < n_{N_{\downarrow \bar \pi}-N_{\downarrow 
(\fm-1)}} 
\leq N_{\downarrow \bar \pi}
\label{Aux A n<n}
\end{eqnarray}

In order to get all quantum numbers which characterize the 
eigenfunction
and obtain the right number of states, one should solve the 
permutation
problem. 
% One can proceed in different manners (e.g. representing the
% permutation chain as a chain of XXX $R$-matrices for specific values 
% of the spectral parameters). 
We will take the result from \cite{vFFR0609} 
(section
on permutation problem of $(gl(\fn|\fm) \oplus gl(2))$ model). We 
get additional sets of integers 
% (quantum number at each step when we eliminate a
% particle as the vacuum): 
\begin{equation} 
1 \leq n^{(k)}_{1}<n^{(k)}_{2}<...<n^{(k)}_{N_{\downarrow (\fq+1)
}+...+N_{\downarrow( k-1)}} \leq N_{\downarrow (\fq+1) 
}+...+N_{\downarrow
k} \,, \quad k= \fq+2, ..., \fm-2\,. 
\end{equation} 
One can remark that if we set $k= \fm-1$ in the above condition, we 
recover
the relation (\ref{Aux A n<n}). This result ends the first subcase.

\paragraph{III-B) A $\bar \pi \uparrow$-particle is on the $j$ site.} 

The periodic boundary conditions on the coefficients 
$\Phi_{\bar{A}}(P Q,
P^{-1})$ can be written analogously to the previous case 
\begin{equation}
\frac{\Phi_{\bar{A}}(P Q C_{K} , P^{-1})}{\Phi_{\bar{A}}(P Q , 
P^{-1})} =
\prod_{l=1}^{N} \sigma_{l}(\lambda_{PQ(K)})
\label{Aux eq:period-glnglm f}
\mb{for} Q \mb{such that} Q(K) \in \cM_{\bar \pi \uparrow}\,.
\end{equation}
 
We proceed as in the case III-A, but instead we 
choose $Q=
Q_{\bar \pi \uparrow} C^{K-N_{ \uparrow \pi}-N_{ \uparrow \bar 
\pi}}_{K} $.
Here $Q_{\bar \pi \uparrow}$ defined similarly to $Q_{\bar 
\pi
\downarrow}$: $Q_{\bar \pi \uparrow}(i)\in \cM_{\bar \pi \uparrow}$ if
$i\in \cM_{\bar \pi \uparrow}$ and $Q_{\bar \pi \uparrow}(i)=i$ for 
the
remaining indices. The cyclic permutation is chosen such that all 
$\pi$ and
$\bar \pi$ down particles are moved to the begining from their 
original
ordering for $Q=id$, that is $C^{K-N_{ \uparrow \pi}-N_{ \uparrow \bar
\pi}}_{K}(i)=(i + N_{ \uparrow \pi}+N_{ \uparrow \bar \pi})_{\mod K}$ 
for any $i$.
% Thus, it is easy to verify that the necessary condition $Q(K) \in 
% \cM_{\bar \pi \uparrow}$ is satisfied. 
This choice implies in particular 
the following relation:
\begin{eqnarray}
PQC_{K} &=& \Big( \prod_{i \in \cM_{\pi \downarrow}}^{\longrightarrow}
\Pi^{i}_{\tilde{P}(N_{ \uparrow \pi}+N_{ \uparrow \bar \pi})} \Big) 
\Big(
\prod_{i \in \cM_{\bar \pi \downarrow}}^{\longrightarrow} 
\Pi^{P_{\bar \pi
\downarrow}(i)}_{\tilde{P}(N_{ \uparrow \pi}+N_{ \uparrow \bar \pi})} 
\Big)
\Big( \prod_{i \in \cM_{\pi \uparrow}}^{\longrightarrow}
\Pi^{i}_{\tilde{P}(N_{ \uparrow \pi}+N_{ \uparrow \bar \pi})} \Big) 
\times
\nonumber \\
&&\times \; \Big( \Pi^{\tilde{P}(N_{\uparrow \pi}+1)}_{\tilde{P}(N_{ 
\uparrow 
\pi}+N_{ \uparrow \bar \pi})} ... \Pi^{\tilde{P}(N_{ \uparrow 
\pi}+N_{ \uparrow \bar \pi}-1)}_{\tilde{P}(N_{ \uparrow \pi}+N_{ 
\uparrow \bar \pi})} \Big) PQ
\end{eqnarray}
where $\tilde{P}=P_{\bar \pi \uparrow}Q_{\bar \pi \uparrow}$ and $PQ =
P_{\bar \pi \downarrow} \tilde{P} C^{K-N_{ \uparrow \pi}-N_{ \uparrow 
\bar
\pi}}_{K}$. Here, due to the choice of $Q$, $P_{\pi \downarrow}$ and $P_{ \pi \uparrow}$ are
 equal to identity. The symbol
$\displaystyle \prod_{i \in \cM_{A}}^{\longrightarrow}
\Pi^{i}_{\tilde{P}(N_{ \uparrow \pi}+N_{ \uparrow \bar \pi})}$ 
denotes the
ordered product of permutation: $...\Pi^{i}_{\tilde{P}(N_{ \uparrow
\pi}+N_{ \uparrow \bar \pi})} \Pi^{i+1}_{\tilde{P}(N_{ \uparrow 
\pi}+N_{
\uparrow \bar \pi})}...$

\medskip

Next, we use relations (\ref{PhiPab=Phi f g}), (\ref{PhiPab=Phi pi up 
pibar
up}) and (\ref{PhiPab=Phi pi down pibar up}) to simplify the 
coefficient
$\Phi^{P Q C_{K}}_{P^{-1}}$:
\begin{eqnarray}
\Phi_{\bar{A}}(P Q C_{K} , P^{-1}) &=&
\prod_{i \in \cM_{\pi \downarrow}} e^{i k_{a_{i}}} 
\sigma_{a_{i}}(\lambda_{\tilde{P}(N_{ \uparrow \pi}+N_{ \uparrow \bar 
\pi})}) 
\prod_{i \in \cM_{\bar \pi \downarrow}} b_{i}^{-1}  
\prod_{i \in \cM_{\pi \uparrow}} e^{-i k_{a_{i}}} 
\sigma_{a_{i}}(\lambda_{\tilde{P}(N_{ \uparrow \pi}+N_{ \uparrow \bar 
\pi})}) \times 
\nonumber \\ 
&&\times \Phi_{\bar{A}} \Big(\Pi^{\tilde{P}(N_{\uparrow 
\pi}+1)}_{\tilde{P}(N_{ \uparrow \pi}+N_{ \uparrow \bar \pi})} ... 
\Pi^{\tilde{P}(N_{ \uparrow \pi}+N_{ \uparrow \bar 
\pi}-1)}_{\tilde{P}(N_{ \uparrow \pi}+N_{ \uparrow \bar \pi})} P Q, 
P^{-1} \Big)
\end{eqnarray}
and finally we get 
\begin{eqnarray}
\prod_{i \in \cM_{\pi \downarrow}} e^{i k_{a_{i}}} 
\sigma_{a_{i}}(\lambda_{\tilde{P}(N_{ \uparrow \pi}+N_{ \uparrow \bar 
\pi})}) 
\prod_{i \in \cM_{\bar \pi \downarrow}} b_{i}^{-1}  
\prod_{i \in \cM_{\pi \uparrow}} e^{-i k_{a_{i}}} 
\sigma_{a_{i}}(\lambda_{\tilde{P}(N_{ \uparrow \pi}+N_{ \uparrow \bar 
\pi})}) \times  
\nonumber \\
\Phi_{\bar{A}} \Big(\Pi^{\tilde{P}(N_{\uparrow \pi}+1)}_{\tilde{P}
(N_{ \uparrow \pi}+N_{ \uparrow \bar \pi})} ... 
\Pi^{\tilde{P}(N_{ \uparrow \pi}+N_{ \uparrow \bar \pi}-1)}_{\tilde{P}
(N_{ \uparrow \pi}+N_{ \uparrow \bar \pi})} P Q, P^{-1} \Big)
\ =\ \prod_{l=1}^{N} \sigma_{l}(\lambda_{\tilde{P}(N_{ \uparrow 
\pi}+N_{ 
\uparrow \bar \pi})}) \Phi_{\bar{A}}(PQ,P^{-1})
\end{eqnarray}

Now we introduce similarly as in previous case the vector 
$\hat{\Phi}_{\bar \pi \uparrow}(P)$:
\begin{equation}
\hat{\Phi}_{\bar \pi \uparrow}(P) \equiv \sum_{\bar{A} \in F} \; 
\sum_{Q'\in \fS_{N_{\bar \pi \uparrow}}} \Phi_{\bar{A}}(P,Q') 
\prod_{i \in \cM_{\bar \pi \uparrow}} e^{\bar{A}_{Q'(i)}}_{i}
\end{equation}
where $F=\spn\{(\fp+1) \uparrow, ..., \fn \uparrow \}$. 
% summation on $\bar{A}\in F$ means that we take $\bar{A}_{i}$ (defined 
% in
% (\ref{excited st aux pr1 glnglm univ})) for $i \in \cM_{\bar \pi 
% \uparrow}$
% equal to all possible values in $F$. The ordering is also chosen such 
% that
% for $Q'=id$, the vector $\prod_{i \in \cM_{\bar \pi \uparrow}}
% e^{\bar{A}_{i}}_{i}$ is equal to $\overbrace{e^{(\fp+1) \uparrow}\otimes
% ....\otimes e^{(\fp+1) \uparrow}}^{N_{\uparrow (\fp+1) }}\otimes ... 
% \otimes
% \overbrace{e^{\fn \uparrow}\otimes ...\otimes e^{\fn
% \uparrow}}^{N_{\uparrow \fn }}$. 
\\
The relations (\ref{PhiPab=Phi f f}) and (\ref{Aux Phi identpart}) 
can be gathered in $\hat{\Phi}_{\bar \pi \uparrow}(P)$:
\begin{equation}
\hat{\Phi}(\Pi_{ab}P)=\alpha^{-1}_{ab} \cP_{ab} \hat{\Phi}(P), \quad 
\text{with} \quad P^{-1}(a)-P^{-1}(b) = -1
\end{equation}
where $\cP_{ab}$ is the permutation acting on the particles situated 
on $a$ and $b$ positions in $\hat{\Phi}_{\bar \pi \uparrow}(P)$. \\
Therefore, the periodicity problem can be rewritten as follows: 
% \begin{eqnarray}
% \prod_{i \in \cM_{\pi \downarrow}} e^{i k_{a_{i}}} 
% \sigma_{a_{i}}(\lambda_{\tilde{P}(N_{ \uparrow \pi}+N_{ \uparrow \bar 
% \pi})}) 
% \prod_{i \in \cM_{\bar \pi \downarrow}} b_{i}^{-1}  
% \prod_{i \in \cM_{\pi \uparrow}} e^{-i k_{a_{i}}} 
% \sigma_{a_{i}}(\lambda_{\tilde{P}(N_{ \uparrow \pi}+N_{ \uparrow \bar 
% \pi})}) \times 
% \nonumber \\
% \hat{\Phi}_{\bar \pi \uparrow} \Big(\Pi^{\tilde{P}
% (N_{\uparrow \pi}+1)}_{\tilde{P}(N_{ \uparrow \pi}+N_{\uparrow \bar 
% \pi})}
% .... \Pi^{\tilde{P}(N_{ \uparrow \pi}+N_{\uparrow \bar 
% \pi}-1)}_{\tilde{P}
% (N_{ \uparrow \pi}+N_{ \uparrow \bar \pi})} \tilde{P} \Big)
% \ =\ \prod_{l=1}^{N} \sigma_{l}(\lambda_{\tilde{P}(N_{ \uparrow 
% \pi}+N_{ 
% \uparrow \bar \pi})}) \hat{\Phi}_{\bar \pi \uparrow}(\tilde{P}) 
% \end{eqnarray}
% which can be simplified into 
\begin{eqnarray} 
X_{m}
\cP_{\tilde{P}(N_{ \uparrow \pi}+ 1) m}...\cP_{\tilde{P}(N_{\uparrow 
\pi}+N_{ \uparrow \bar \pi}-1) m}  \hat{\Phi}_{\bar \pi 
\uparrow}(\tilde{P}) = \prod_{l=1}^{N} \sigma_{l}(\lambda_{m}) 
\hat{\Phi}_{\bar \pi \uparrow}(\tilde{P}),
\end{eqnarray}
with $m=\tilde{P}(N_{ \uparrow \pi}+N_{ \uparrow \bar \pi})$ and 
\begin{equation}X_{m}=
\prod_{l\in \cM_{\bar \pi \uparrow}, l\neq m} \alpha_{ml} 
\prod_{i \in \cM_{\pi \downarrow}} e^{i k_{a_{i}}} 
\sigma_{a_{i}}(\lambda_{m}) 
\prod_{i \in \cM_{\bar \pi \downarrow}} b_{i}^{-1}  
\prod_{i \in \cM_{\pi \uparrow}} e^{-i k_{a_{i}}} 
\sigma_{a_{i}}(\lambda_{m})
\end{equation}
Choosing $\tilde{P} = C^{-m}_{\bar \pi \uparrow}$ 
with $C_{\bar \pi \uparrow}=\Pi^{N_{ \uparrow \pi} + 1}_{N_{ \uparrow 
\pi}+N_{ \uparrow \bar \pi}}...\Pi^{N_{ \uparrow \pi}+N_{ \uparrow 
\bar \pi}-1}_{N_{ \uparrow \pi}+N_{ \uparrow \bar \pi}}$, we 
find the Bethe equations:
\begin{eqnarray}
&&\prod_{l\in \cM_{\bar \pi \uparrow}, l\neq m} \!\! \alpha_{ml} 
\prod_{i \in \cM_{\pi \downarrow}} e^{i k_{a_{i}}} 
\sigma_{a_{i}}(\lambda_{m}) 
\prod_{i \in \cM_{\bar \pi \downarrow}} b_{i}^{-1} 
\prod_{i \in \cM_{\pi \uparrow}} e^{-i k_{a_{i}}} 
\sigma_{a_{i}}(\lambda_{m})
\, e^{\frac{2 \pi i}{N_{\uparrow \bar \pi}} \sum_{i=1}^{N_{ \uparrow 
\bar \pi}-N_{\uparrow \fn}} \bar n_{i}} = \prod_{l=1}^{N} 
\sigma_{l}(\lambda_{m})
\nonumber \\
&&\mb{with} m \in \cM_{\uparrow \bar \pi}\mb{and}
1 \leq \bar n_{1} < ... < \bar n_{N_{\uparrow \bar \pi}-N_{\uparrow 
\fn}} \leq N_{\uparrow \bar \pi}
\label{Aux B n<n}
\end{eqnarray}

The problem of the complete characterization of the eigenfunction and 
of
state counting arises in a similar way as in case III-A. We get 
additional sets of integers:
\begin{equation}
 1 \leq  n^{(k)}_{1}<n^{(k)}_{2}<...<n^{(k)}_{N_{\uparrow (\fp+1) 
}+...+N_{\uparrow (k-1)}} 
\leq  N_{\uparrow (\fp+1) }+...+N_{\uparrow k} \,, \quad
 k= \fp+2, ..., \fn-1\,.\
\end{equation}
Again, we recover relation (\ref{Aux B n<n}) if we 
take $k= \fn$ in the above condition.


\begin{thebibliography}{99}

\bibitem{Hubbard}
J.~Hubbard, \textsl{Electron Correlations in Narrow Energy Bands}, 
Proc. Roy. Soc. London  \textbf{A276} (1963) 238; 
\\
\textsl{Electron Correlations
in Narrow Energy Bands II. The Degenerate Band Case}, \textit{ibid.}
\textbf{A277} (1964) 237.

\bibitem{Gutzwiller}
M.C.~Gutzwiller, \textsl{Effect of Correlation on the Ferromagnetism 
of Transition Metals}, Phys. Rev. \textbf{10} (1963) 15.

\bibitem{Monto}
A.~Montorsi, \textsl{The Hubbard Model}, World Scientific Singapore, 
(1992).

\bibitem{book} F.~E{\ss}ler, H.~Frahm, F.~Goehmann, A.~Klumper and
V.~Korepin, \textsl{The One-Dimensional Hubbard Model},
Cambridge University Press 2005.

\bibitem{LiebWu}
E.H.~Lieb and F.Y.~Wu, \textsl{Absence of Mott transition in an exact
solution of the short-range one-band model in one dimension}, Phys. 
Rev. Lett. \textbf{20} (1968) 1445; 
\\
Erratum, \textit{ibid.} \textbf{21} (1968) 192; 
\\
\textsl{The one-dimensional Hubbard model: a reminiscence}, 
Physica \textbf{A321} (2003) 1 and \texttt{cond-mat/0207529}.

\bibitem{shastry}
B.S.~Shastry, \textsl{Infinite conservation laws in the 
one-dimensional
Hubbard model}, Phys. Rev. Lett. \textbf{56} (1986) 1529; 
\\
\textsl{Exact
integrability of the one-dimensional Hubbard model}, \textit{ibid.}
\textbf{56} (1986) 2453; 
\\
\textsl{Decorated star triangle relations 
and exact
integrability of the one-dimensional Hubbard model}, J. Stat. Phys.
\textbf{50} (1988) 57.

\bibitem{Akutsu}
E.~Olmedilla, M.~Wadati and Y.~Akutsu, \textsl{Yang-Baxter Relations 
for Spin Models and Fermion Models}, J. Phys. Soc. Japan \textbf{56} 
(1987) 2298.

\bibitem{shiro2}
M.~Shiroishi and M.~Wadati, \textsl{Yang-Baxter equation for the 
R-matrix
of the one-dimensional Hubbard model}, J. Phys. Soc. Japan \textbf{64}
(1995) 57.

\bibitem{maasa2} 
Z.~Maassarani, \textsl{The $su(N)$ Hubbard model}, Phys. Lett. 
\textbf{A239} (1998) 187, \texttt{cond-mat/9709252}; \\
\textsl{Exact
integrability of the $su(N)$ Hubbard model}, Mod. Phys. Lett. 
\textbf{B12} (1998) 51, \texttt{cond-mat/9710083}.

\bibitem{Rej:2005qt}
A.~Rej, D.~Serban and M.~Staudacher, \textsl{Planar N=4 gauge theory 
and the Hubbard model}, JHEP \textbf{0603} (2006) 018 and 
\texttt{hep-th/0512077}.

\bibitem{XX} J. Drummond, G. Feverati, L. Frappat and E. Ragoucy,
\textsl{Super-Hubbard models and applications}, 
JHEP \textbf{0705} (2007) 05008 and \texttt{hep-th/0703078}.

\bibitem{FFR}
G. Feverati, L. Frappat and E. Ragoucy,
\textsl{Universal Hubbard models with arbitrary symmetry,} 
JSTAT \textbf{0904} (2009) P04014 and \texttt{arXiv:0903.0190 
[math-ph]}.

\bibitem{vFFR0609}
V. Fomin, L. Frappat and E. Ragoucy,
\textsl{Bethe equations for generalized Hubbard models,}
JHEP \textbf{0909} (2009) 055 and \texttt{arXiv:0906.4512}.

\bibitem{phasestring} 
G. Arutyunov, S. Frolov, M. Staudacher,
\textit{Bethe Ansatz for Quantum Strings},
 JHEP \textbf{0410} (2004) 016, \texttt{arXiv:hep-th/0406256}.

\bibitem{string}
T. Bargheer, N. Beisert, F. Loebbert,
\textit{Long-Range Deformations for Integrable Spin Chains},
J. Phys. \textbf{A42} (2009) 285205,
\texttt{arXiv:0902.0956}

\bibitem{Takahashi72} M. Takahashi, \textit{One-dimensional Hubbard 
model at finite temperature}, Prog. Theor. Phys. \textbf{47} (1972) 
69;\\  
T. Deguchi, F. H. L. Essler, F. Gohmann, A. 
Klumper, V. E. Korepin, and K. Kusakabe. 
\textit{Thermodynamics and excitations of the one-dimensional Hubbard 
model}, Phys. Rep. \textbf{331} (2000)
197-281  and \texttt{arXiv:cond-mat/9904398}.

\bibitem{dolcini} F. Dolcini and A. Montorsi,
\textit{Results on the symmetries of integrable fermionic 
models on chains}, Nucl. Phys. \textbf{B592} (2001) 563
and \texttt{arXiv:cond-mat/0110246}.

\bibitem{sutherland} B. Sutherland, \textit{An introduction to the 
Bethe ansatz},
Lect. Note Phys. \textbf{242}, eds B. 
Shastry, S. Jha and V. Singh, Springer (1985) Berlin.

\end{thebibliography}
\end{document}